\documentclass[%
 reprint,
superscriptaddress,
showpacs,
amsmath,amssymb,
prb,
floatfix,
]{revtex4-1}
\usepackage{graphicx}% Include figure files
\usepackage{subfigure} % use for side-by-side figures
\usepackage[colorlinks,linkcolor=blue,anchorcolor=green,citecolor=blue]{hyperref}
\usepackage{bm}%
\usepackage[inline]{enumitem}
\usepackage{footnote}
\usepackage{multirow}

\begin{document}

 \title{Tailoring the electrocaloric effect by internal bias fields and field protocols}% Force line breaks
 
 \author{Yang-Bin Ma}
 \affiliation{Institute of Materials Science, Technische Universit\"at Darmstadt\ \\ 64287 Darmstadt, Germany}
 \author{Bai-Xiang Xu}
\email{xu@mfm.tu-darmstadt.de}
\affiliation{Institute of Materials Science, Technische Universit\"at Darmstadt\ \\ 64287 Darmstadt, Germany}
 \author{Karsten Albe}
\affiliation{Institute of Materials Science, Technische Universit\"at Darmstadt\ \\ 64287 Darmstadt, Germany}
 \author{Anna Gr\"{u}nebohm}
 \affiliation{Faculty of Physics and Center for Nanointegration (CENIDE), University of Duisburg-Essen\ \\ 47048 Duisburg, Germany}

\begin{abstract} 
In acceptor doped ferroelectrics and in ferroelectric films and nanocomposites, defect dipoles, strain gradients, and the electric boundary conditions at interfaces and surfaces often impose internal bias fields.
In this work we delicately study the impact of internal bias fields on the 
 electrocaloric effect (ECE), utilizing an analytical model and \emph{ab initio}-based molecular dynamics simulations.
We reveal the complex dependency of the ECE on field protocol and relative strength of internal and external fields. 
The internal fields may even reverse the sign of the response (inverse or negative ECE). 
 We explore the transition between conventional and inverse ECE and discuss reversible and irreversible contributions to the field-induced specific entropy change.
Most importantly, we predict design routes to optimize the cooling and heating response for small external fields by the combination of internal field strengths and the field loading protocol.
%We confirm our predictions by {\it{ab initio}} based molecular dynamics simulations.

\end{abstract}

\date{\today}% It is always \today, today,
  % but any date may be explicitly specified

%\keywords{Suggested keywords}%Use showkeys class option if keyword
    %display desired
\maketitle

\newcommand{\AG}[1]{{\bf **AG: #1**}}

\newcommand{\ma}[1]{\textcolor{red}{\bf **ma: #1**}}

\newcommand{\maa}[1]{\textcolor{cyan}{\bf **ma: #1**}}

\section{Introduction}
The electrocaloric effect (ECE) in ferroelectrics is promising for solid-state cooling devices~\cite{2007_Scott,2014_Moya}
as an external electric field can induce large changes of specific entropy ($\Delta S$) respectively of temperature ($\Delta T$).
Commonly, an external field reduces the specific dipolar entropy and the material heats up under adiabatic field application (conventional or positive ECE). In some cases also adiabatic cooling under field application has been found (inverse or negative ECE).~\cite{Wang2013a,Uddin2013,Axelsson2013,Thacher1968,2017_Marathe,2017Wu,2018Qi} 
The inverse ECE might have the potential to enhance the overall caloric response and attracts considerable research interest, as reviewed in Ref.~\onlinecite{2015_Kutnjak,Liu2016}.

One prominent way to optimize ferroelectric properties is acceptor doping, i.e., the substitution with ions of lower valence.
In this case, charge neutrality is achieved by oxygen vacancies. 
For example, Ti ions in BaTiO$_3$ can be substituted with Mn, Cu, or Fe ions. 
With time, the dopants and oxygen vacancies form associate defect dipoles.~\cite{1978_Jonker} 
As shown experimentally, these dipoles can be unidirectionally aligned, for example, by a long-time poling process.~\cite{2008_Zhang,2010_Folkman} 
Due to the high activation barrier for oxygen vacancy migration, these defect dipoles are almost non-switchable in the ferroelectric phase around room temperature on time-scales relevant for cooling devices.~\cite{2004_Ren,2013_Erhart} 
There are numerous studies dealing with the influence of defect dipoles on material properties 
(see the review article~\onlinecite{2015_Genenko}).
Particularly, it has been observed~\cite{Han2017} that defect dipoles induce internal electrical field which can shift the 
hysteresis, see Fig.~\ref{fig:figure1}. The occurrence and direction of the shift depend on the alignment of the defect dipoles and thus on the previous field and heat treatment of the material.
%In addition, pinched double hysteresis loops have been found if the defect dipoles are aligned with the polarization in domains with opposite directions.~\cite{2014_Tan,2014_Ichikawa}
Large modifications of the functional responses by such doping are possible.\cite{2017Chapman,2004_Ren}

So far, only few studies have explored the impact of acceptor doping on the ECE.
It has been reported that co-doping with Mn and Y modifies the ECE in Ba$_{0.67}$Sr$_{0.33}$TiO$_3$.~\cite{Xu2017}
And our previous studies on doped BaTiO$_3$ revealed the possibility to shift the $\Delta T$-peak to higher temperatures.~\cite{Grunebohm2015,2016_Mab} %, see also Fig.~\ref{fig:para} in Appendix~\ref{sec:fieldECE}. 
Strikingly, we found an inverse ECE in the presence of defect dipoles aligned anti-parallel to the external field and a transition between inverse and conventional ECE with the strength of the external field strength.\cite{Grunebohm2015,2016_Mab}
In addition, we found that the ECE for field application may exceed the ECE for field removal~\cite{2016_Mab} and that the response for different initial states (either prepared by field cooling or field heating) may differ by one order of magnitude in certain temperature intervals.\cite{Grunebohm2015}

Analogous, internal bias fields are commonly observed in nanocomposites and ferroelectric films.\cite{2015_Qian,Qian2016a,Lee2014,Chu2009,Zhou2005}
These bias fields have been related to stress gradients,\cite{Lee2014} asymmetric electric boundary conditions,\cite{Chu2009} and ferroelectric blends in a relaxor matrix. \cite{2015_Qian,Qian2016a}
Qian {\em et al.} reported that the ECE in a relaxor ferroelectric polymer can be enhanced by 45\% when ferroelectric polymer blends are introduced, inducing internal bias fields.~\cite{2015_Qian,Qian2016a}
Similar to our findings for defects, they found inverse ECE in the presence of internal bias fields as well.
Therefore, it is important to obtain a detailed understanding of the impact of internal bias fields on the ECE, and tap the full potentials of the bias fields.

In literature, the ECE is mostly determined indirectly.\cite{2008_Akcay,2011_Dunne,Mangeri2016,Xu2017,Rose2012,2017Wu,Bin-Omran2016}
%The adiabatic temperature change for a field change from $E_\mathrm{init}$ to $E_\mathrm{end}$ is commonly determined \cite{2008_Akcay,2011_Dunne,Mangeri2016,Xu2017} from the thermodynamic Maxwell relation
%\begin{eqnarray} \label{eq:max}
%\Delta T = - \int_{E_\mathrm{init}}^{E_\mathrm{end}} \frac{T}{C(E,T)} \left. \frac{\partial P(E,T)}{\partial T} \right|_E d E,
%\Phi = \Phi_0 + \frac{1}{2} x_2 \eta^2 + \frac{1}{4} x_4 \eta^4 + ... - h \eta.
%\end{eqnarray}
%with the heat capacity $C(E,T)$.
%Alternatively, correlation functions\cite{Bin-Omran2016} have been used to predict entropy and temperature changes. 
However, there are several limitations of the indirect approach as reviewed in Ref.~\onlinecite{2016_Liu}. In particular, irreversible specific entropy contributions related to ferroelectric switching or first order phase transitions are commonly not taken into account.
Alternatively, $\Delta T$ can be measured or simulated directly.~\cite{Wang2012,Novak2013a,Bai2012,2012_Ponomareva,2013_Nishimatsu,2016_Marathe,2015_Maa,2016_Mab,2016_Ma} 
% can be measured directly, e.g., by using thermocouples and differential scanning calorimeters, or simulated, e.g., by using molecular dynamics (MD) or Monte Carlo (MC) simulations.
%Besides the temperature change, also the entropy change plays a significant role in understanding the EC effect. 
%In MD or MC simulations, the entropies can be calculated, which requires a considerable effort.~\cite{2012_Ponomareva,2012_Rose,2013_Nishimatsu,Grunebohm2015,2015_Maa,2016_Mab,2016_Ma} 
So far, direct temperature changes have been mainly determined for a unipolar field loading (field application or removal for one field direction) and the possible impacts of thermal and field hysteresis have been widely neglected.
However, simulations \cite{Marathe2018} and direct measurements~\cite{Moya2013,Bai2013} revealed the impact of thermal hysteresis and field protocol on the ECE close to phase transitions. 
Furthermore, field hysteresis and field direction may strongly influence the ECE in the ferroelectric phase.
% irreversible contributions may play an important role, due to thermal and field hysteresis %and the field loops
%~\cite{Bolten2000}. % or the presence of defect dipoles~\cite{2016_Mab}.
% However, this is far from being well understood.
For instance, based on an analytical model in Ref.~\onlinecite{2016_Mac} it has been found that the overall cooling can be enhanced by field reversal. 
This concept has been verified by experiments~\cite{Thacher1968,Wang2013a,Basso2014,2016_Ma,2016_Mac} and it has been shown that ferroelectric switching and related work losses may result in large modifications of $\Delta T$.
However, we are not aware of a delicate study on the impact of the internal bias fields and different field protocols on the ECE.

%%%%%%%%%%%entropy analysis

%%%%%%%%%%%
In the present paper we combine phenomenological Landau theory\cite{Pirc2011,2016_Mab} with \emph{ab initio}-based MD simulation~\cite{2008_Nishimatsu} in order to shed light on the interplay between external and internal field strengths, field protocols and thermal history in ferroelectrics with internal bias fields. Thereby, we consider reversible and irreversible specific entropy changes.
In particular we predict routes to optimize the ECE for a fixed small magnitude of the applied field, which is important for devices where Joule heating and electrical breakdown limit the strength of the external field strength.

%In the following, we investigate the EC response, taking the irreversible entropy changes during the polarization switching into account in the ferroelectric phase.
%In addition to direct MD simulations,~\cite{2008_Nishimatsu} we extend the approach by Pirc {\em et al.} who proposed an analytical expression for reversible polarization changes based on phenomenological Landau theory.~\cite{Pirc2011} Previously, in the analytical model for defect-free samples, instead of assuming constant total entropy with purely reversible polarization changes,~\cite{Pirc2011} we have considered changes of the total entropy with irreversible polarization changes, and predicted an enhancement of the EC cooling by applying a proper reversed field, which has been verified by experiments.~\cite{2016_Mac,2016_Mab} 

The paper is organized as follows. First, the used model is elaborated in Sec.~\ref{sec:model}. Second, the influence of internal bias fields on the ECE is discussed in Subsec.~\ref{subsec:density}.
Hereby, we focus on the impact of different strengths of the internal field which is anti-parallel to the external field. 
Supplementarily, the impact of the external field strength for a fixed internal field and the trends for parallel external and internal fields are discussed in Appendixes ~\ref{sec:fieldECE} and \ref{sec:para}, respectively.
We confirm our predictions for the example of defect dipoles by {\it ab initio}-based molecular dynamics simulations in Appendix.~\ref{subsec:md}.
Based on the obtained knowledge, we predict how the ECE in the presence of internal bias fields can be optimized by the field protocol in Subsec.~\ref{subsec:cycle}. 
Finally, conclusions and outlook can be found in Sec.~\ref{sec:conclusion}.

\section{analytical Model}
\label{sec:model}
\begin{figure} 
	\centering 
	\centerline{\includegraphics[height=0.3\textwidth]{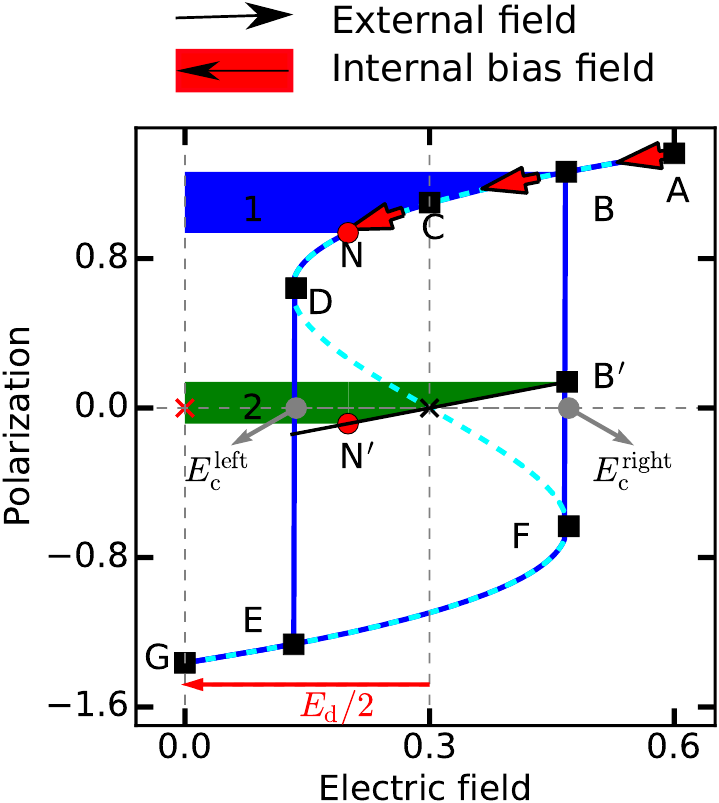}}
	\caption{Schematic hysteresis loop in the presence of internal fields induced by anti-parallel defect dipoles or nanocomposite blends. Several characteristic points X with $(E_\mathrm{X} , P_\mathrm{X})$ can be defined:
D and F are the inflection points, and C corresponds to the field center of the hysteresis which is shifted from the point of origin to $
E_d/2$.
		Between A--B and E--G the polarization change is reversible. For intermediate field strengths irreversible changes of the polarization are induced and the reversible contribution to the polarization, $P_\mathrm{r}$, is given by the black line, which crosses the center of hysteresis.
		Blue and green areas correspond to the actual work density and its reversible part, for a reduction of the external field from $E_\mathrm{B}$ to $E_\mathrm{N}$, see text.
		\label{fig:figure1}	}
\end{figure}
%\begin{figure} [htp]
%	\centering %
%	\centerline{\includegraphics[height=0.3\textwidth]{fig2.pdf}}
%	\caption{Graphic illustration of the work loss if the external field changes from B to N.
%The green area, corresponding to reversible work in Fig.~\ref{fig:figure1} is shifted by (0, $P_\mathrm{B}-P_\mathrm{B^\prime}$). The gray area represent the negative work loss (actual work - reversible work). \AG{Wouldn't it make sense to add the pink loss for field on here, too?}}
%	\label{fig:figure2}
%\end{figure}
Pirc {\em et al.} proposed an analytical model to determine the ECE for reversible changes of the polarization based on Landau theory.~\cite{Pirc2011} In Ref.~\onlinecite{2016_Mac} irreversible changes of the polarization in the course of ferroelectric switching have been added to this model. 
Internal bias fields can be included in the Landau model straightforwardly.
For example, fixed defect dipoles induce a polarization $P_\mathrm{d}$ parallel to the polarization of their surrounding during equilibration, which couples with the polarization of the free dipoles $P$ and gives rise to the internal electric field $E_\mathrm{d}$. % = 2J P_\mathrm{d}$.
We assume a coupling $E_\mathrm{d} = 2J P_\mathrm{d}$ with $J=3.0$.
One can define the mean field free energy density $F_\mathrm{dip}$ as
\begin{eqnarray} \label{eq:mean}
F_\mathrm{dip} = F_0 + \frac{1}{2} a P^2 + \frac{1}{4} b P^4 - E P - \frac{1}{2} E_\mathrm{d} P,
\end{eqnarray}
with $E$ the external field. According to Kutnjak {\em et al.}, we use dimensionless quantities ($T_0= 1$, $ b = 1/3 $, and $ a_0 = \partial a / \partial T = 1$),~\cite{2015_Kutnjak} neglect the temperature dependency of $b$ and $E_\text{d}$, and assume
 $a = a_0 (T - T_0) $ with $T$ and $T_0$ the temperature and the Curie temperature, respectively.

At equilibrium, the free energy density has a local extremum ($\partial F_\mathrm{dip} / \partial P=0$) resulting in 
% % % % % actual field
\begin{eqnarray} 
{ E } &=& a P + b { P }^3 - E_\mathrm{d}/2.
\label{eq:act}
\end{eqnarray}
In the ferroelectric phase ($T<T_0$) the equilibrium P(E) curve is thus S-shaped and between the left and right coercive fields (points E and B in Fig.~\ref{fig:figure1}) multiple metastable states exist. 
In this hysteretic region, changes of the external field induce reversible and irreversible changes of $P$. 
As shown experimentally by Bolten {\em et al.},~\cite{2000_Bolten,2003_Bolten} the reversible changes of $P$ are given by the straight line passing through the center of the hysteresis with the slope of E(P) at the coercive field
 ($ \partial E/\partial P \mid _ {B} = a + 3 b {P_\mathrm{B}} ^2 $).
  %, where $ E_\mathrm{B} = a P_\mathrm{B} + b P_\mathrm{B}^3 - J P_\mathrm{d}$) 
%For field strengths between points B and E, the multiple local extrema of $F$ result in irreversible field induced changes of $P$.
%The applied field does the work $W_\mathrm{actual}=\int_{P_\mathrm{B}}^{P} E dP$ and thus irreversible changes of $P$ induce work losses.
%Between point A and B, the polarization change is reversible with $W_\mathrm{loss}=0$ and the change in the dipolar entropy in Eq.~(\ref{eq:deltasdip}) determines the final temperature.
%% % % % % % % 
%% % % % % % % 
%% % % % % % % reversible process
 % % % % % % field of reversible process
The reversible polarization $P_{\mathrm{r}}$ and the corresponding field $E$ thus satisfy the relation
%\begin{eqnarray} 
%E + E_\mathrm{d}/2 = (a + 3 b {P_\mathrm{B}} ^2 ) P_{\mathrm{r}}, \nonumber
%\end{eqnarray} 
%or 
\begin{eqnarray} 
E = a P_{\mathrm{r}} + 3 b {P_\mathrm{B}}^2 P_{\mathrm{r}} - E_\mathrm{d}/2 .
\label{eq:rev}
\end{eqnarray}

% % % % % work lost
By substituting Eq.~(\ref{eq:act}) into (\ref{eq:rev}), one obtains 
%\begin{eqnarray} \label{eq:work}
%{ a P + b P ^3 - J P_\mathrm{d} } &=& a P_{\mathrm{r}} + 3 b { P_{ \mathrm{B} } } ^2 P_{\mathrm{r}} - J P_\mathrm{d}. \nonumber
%\end{eqnarray}
the reversible polarization: 
\begin{eqnarray}
{ P_{\mathrm{r}} } &=& 
\frac{ a P + b P^3 }{ a + 3 b { P_{ \mathrm{B} } } ^2 }. \nonumber
\end{eqnarray}

% % % % % % reversible polarization

If the field is varied between $E_\mathrm{init}$ and $E_\mathrm{end}$, it induces a change of $P$, which is related to the 
work density $W_\mathrm{actual}=\int_{P_\mathrm{init}}^{P_\mathrm{end}} E dP$, with $P_\mathrm{init}$ and $P_\mathrm{end}$ the initial and final polarization, respectively.
For positive $E$, the work is thus positive if $|P|$ increases, while negative if $|P|$ decreases, and vice versa for negative $E$.
Outside the hysteretic region (e.g.\ between A and B or E and G in Fig.~\ref{fig:figure1}), the field-induced work is fully reversible. In the hysteretic region (e.g.\ between B and E in Fig.~\ref{fig:figure1}), one can divide the actual work density into the reversible work density done on P$_r$ and irreversible work loss density. Without switching of the polarization direction (e.g.\ between B and D in Fig.~\ref{fig:figure1}) the losses are given as % % % % % % work lost
\begin{eqnarray} 
{ W_\mathrm{loss}} &=& W_\mathrm{actual} - W_\mathrm{r} \nonumber \\
&=& \int_{P_\mathrm{B}}^{P} E dP - \int_{P_{\mathrm{B}^\prime}}^{P_\mathrm{r}} E dP_{\mathrm{r}} \nonumber \\
&=& \frac{1}{2} a { P }^2 + \frac{1}{4} b { P }^4 
- ( \frac{1}{2} a { P_{\mathrm{B}} }^2 + \frac{1}{4} b { P_{\mathrm{B}} }^4 ) + \frac{1}{2} a { P_{\mathrm{B^\prime}} }^2 \nonumber \\ 
&+& \frac{3}{2} b { P_{\mathrm{B}} }^2 { P_{\mathrm{B^\prime}} }^2 
- ( \frac{1}{2} a { P_\mathrm{r} }^2 + \frac{3}{2} b { P_{\mathrm{B}} }^2 { P_\mathrm{r} }^2 ) \nonumber \\
&-& ( P - P_{\mathrm{B}} - P_\mathrm{r} + P_{\mathrm{B}^\prime}) E_\mathrm{d}/2 ,
\label{eq:lost}
\end{eqnarray}
where $P_\mathrm{B^\prime}=\left.P_\mathrm{r}\right|_{P=P_{\mathrm{B}}} = \dfrac{ a P_\mathrm{B} + b P_\mathrm{B}^3 }{ a + 3 b { P_{ \mathrm{B} } } ^2 }$.
%which represents the hysteretic path and 
 In the course of switching (e.g., between D and E in Fig.~\ref{fig:figure1}), the losses are furthermore enhanced by $(P- P_{\mathrm{D}}) E_{\mathrm{D}}$.
 The actual and reversible work density for a field variation from B to N are 
 illustrated in Fig.~\ref{fig:figure1} in blue and green, respectively. 

%In summary, the loss is given as
%\begin{equation} 
%{W_\mathrm{loss}} = 
%\begin{cases}
%0 & \text{between A and B },
%\\
%W_\mathrm{loss1} 
%& \text{between B and D },
%\\
%W_\mathrm{loss1}+ (P - P_{\mathrm{D}}) E_{\mathrm{D}}
%& \text{between D and E },
%\\
%0 & \text{between E and G }.
%\nonumber
%\end{cases}
%\end{equation}

 %Between point B and E the free energy has multiple local extrema, see Eq.~(\ref{eq:mean}), and losses must be taken into account, i.e., $\Delta S_\mathrm{total} \neq 0$. In this case, the change of total entropy can be calculated from the work loss $W_\mathrm{loss}$ by
%If the green area is shifted by (0, $P_\mathrm{B}-P_\mathrm{B^\prime}$), the remaining part of the blue area given in gray in Fig.~\ref{fig:figure2} illustrates the corresponding work losses.

Depending on the conditions (isothermal, adiabatic or mixed), a variation of the external field induces changes of the specific entropy and/or temperature of the system.
In order to determine the adiabatic temperature change, it is convenient to separate the specific entropy change into the field-induced reversible ($\Delta S_\mathrm{dip}$) and irreversible ($\Delta S _ \mathrm{total} = \frac{ W_\mathrm{loss}}{T_\mathrm{init}}$) changes of the dipolar degrees of freedom
% % % % the change of $S_\mathrm{dip}$ from the initial state ($P_\mathrm{init}$, $E_\mathrm{init}$, $T_\mathrm{init}$) to final state ($P$, $E$, $T$) is given as
, and the remaining vibrational degrees of freedom ($S_\mathrm{vib}$) which depend on the external field only weakly \cite{Pirc2011,2015_Kutnjak,2016_Mac}:
% % % % % % vibrational entropy
\begin{eqnarray} \label{eq:T-Sdip}
\Delta S_\mathrm{vib}&=&\Delta S_\mathrm{total}-\Delta S _ \mathrm{dip}.
\end{eqnarray} 
$S_\mathrm{dip}$ is given as 
\begin{eqnarray} \label{eq:deltasdip}
{ \Delta S_\mathrm{dip} } &=&
S_\mathrm{dip}(P) - S_\mathrm{dip}(P_\mathrm{init}) \nonumber
\\
&=& - \frac{1}{2} a_0 (P^2 - {P_\mathrm{init}}^2)\;.
\end{eqnarray}
with $S_\mathrm{dip} = - \partial F_\mathrm{dip} /\partial T = - \frac{1}{2} a_0 P^2 $.

Neglecting the weak temperature-dependency of the specific heat capacity of the non-polar degrees of freedom,\footnote{We note that the temperature dependency of $c_\mathrm{ph}$ in the temperature range of interest is negligible apart from phase transitions. 
} $c_\mathrm{ph}$ is taken as 15 in the reduced units according to Ref.~\onlinecite{Pirc2011}.
The change of the specific vibrational entropy is given by the initial $T_\mathrm{init}$ and final temperature $T_\mathrm{end}$ as 
\begin{eqnarray} \label{eq:T}
\Delta S_\mathrm{vib} 
= \int_{T_\mathrm{init}}^{T_\mathrm{end}} \frac{c_\mathrm{ph}}{T} dT \cong c_\mathrm{ph} \ln(T_\mathrm{end} / T_\mathrm{init}), 
\end{eqnarray}
%&\cong& -c_\mathrm{ph} \ln(T_\mathrm{final} / T_\mathrm{init}),
 %\Delta S_\mathrm{dip}&=& S_\mathrm{dip}(P) - S_\mathrm{dip}(P_\mathrm{init}) 
and the adiabatic temperature change can be determined by
\begin{eqnarray} \label{eq:sol2}
\Delta T=T_\mathrm{init}\exp \big[ ( W_\mathrm{loss} / T_\mathrm{init} - \Delta S_\mathrm{dip} ) / c_\mathrm{ph} \big] - T_\mathrm{init} .
\end{eqnarray}
%\AG{What do you think about this rearrangement? If we keep it, we have to check the references to the different equations}

\begin{figure} [t]
	\centering 
	\centerline{\includegraphics[height=0.4\textwidth]{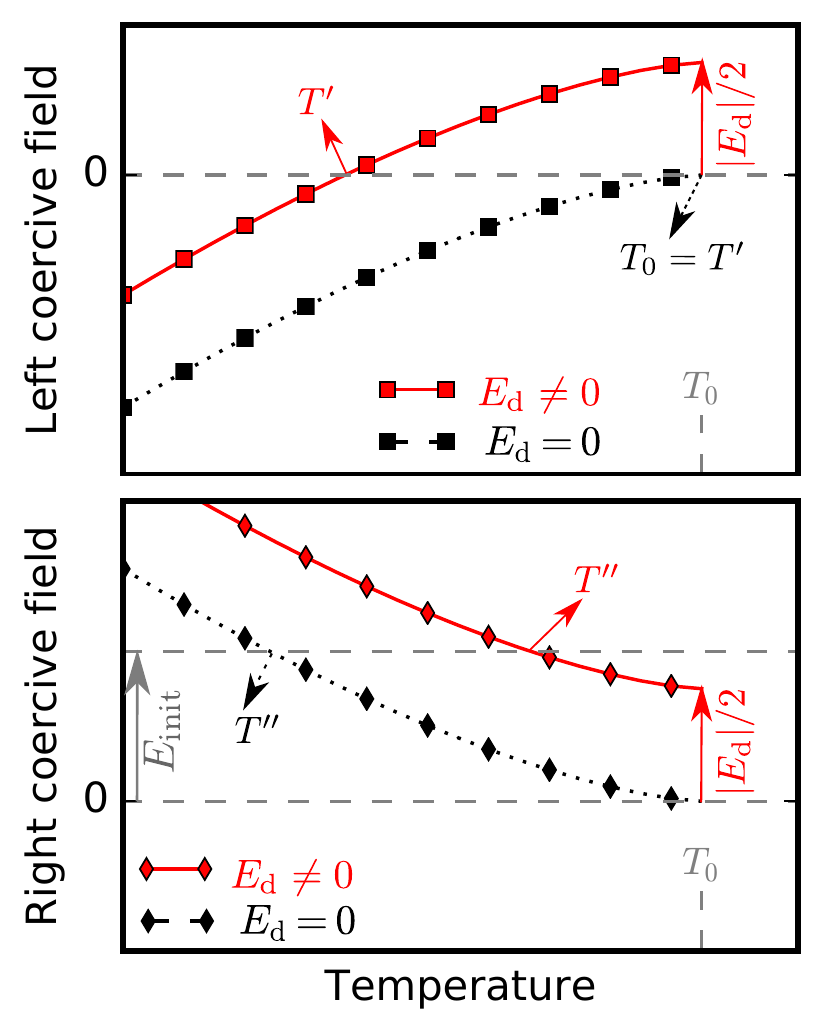}}
	\caption{$T'$ and $T''$, determined by the external fields $E_\mathrm{init}$ and the variation of the left and right coercive field ($E_\mathrm{c}^\mathrm{left}$, $E_\mathrm{c}^\mathrm{right}$), see Fig.~\ref{fig:figure1}. $T'$ indicates the temperature at which $E_\mathrm{c}^\mathrm{left}=0$, without bias fields (dashed lines with black squares) and for induced bias field $E_\mathrm{d}$ (solid lines with red squares). $T''$ indicates the temperature at which $E_\mathrm{c}^\mathrm{left}=E_\mathrm{init}$, without bias fields (dashed lines with black diamonds) and for induced bias field $E_\mathrm{d}$ (solid lines with red diamonds).
	}
	\label{fig:figure3}
\end{figure}

% without defects
\begin{figure*}[t]
	\centering 
	\centerline{\includegraphics[height=0.4\textwidth]{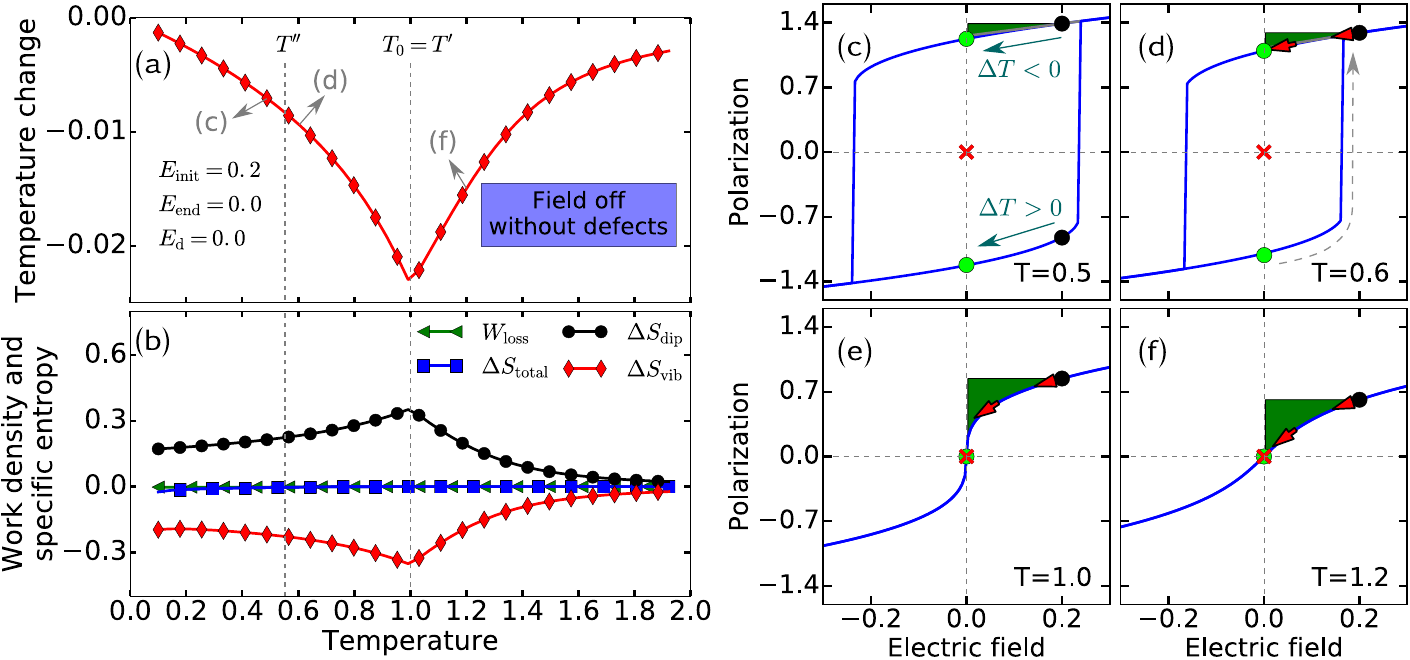}}
	\caption{ \textbf{Case (i):} Electrocaloric effect without internal bias fields in the sample with positively poled initial states.
	Subfigures illustrate (a) the temperature change $\Delta T$, (b) the corresponding work loss density and specific entropy changes and (c)-(f) representative hysteresis loops. In (c)-(f) dots, crosses and arrows illustrate initial and final state, the center of the hysteresis, and the direction of the field change. Ramping off the field, (c) demonstrates the conventional ECE on the upper branch, and the inverse ECE on the lower branch.
%	There meanings of symbols are also applied in all the following figures. 	
%Losses are negligible for all temperatures and the reversible dipolar entropy change is maximal around the Curie temperature.
	\label{fig:defect1}}
\end{figure*}

For the discussion of the temperature-dependency of the ECE, it is convenient to define two characteristic temperatures for each combination of external and internal field. %$T'$ and $T''$ for which $E_\mathrm{c}^\mathrm{left}=0$ and $E_\mathrm{c}^\mathrm{right}=E$.
At low temperatures, the field hysteresis is broad and thus $E_\mathrm{c}^\mathrm{left}$ is negative and $E_\mathrm{c}^\mathrm{right}$ is large and potentially exceeds the applied field $E$.
With increasing temperature, the hysteresis is systematically reduced i.e., $E_\mathrm{c}^\mathrm{left}$ and $E_\mathrm{c}^\mathrm{right}$ increase and decrease, respectively.
At $T'$ the left coercive field changes sign, and at $T''$ the external field exceeds $E_\mathrm{c}^\mathrm{right}$ (see Fig.~\ref{fig:figure3}). 
The polarization at the coercive fields is given as $P_\mathrm{D,F} = \pm \sqrt{- a/3b }$.
 According to Eq.~\eqref{eq:act} the left and right coercive fields are thus given as
\begin{eqnarray}\label{eq:Ecleft}
E_\mathrm{c}^\mathrm{left} &=& \frac{2}{3} a \sqrt{-\frac{a}{3b}} - E_\mathrm{d}/2\;, \\
E_\mathrm{c}^\mathrm{right} &=& -\frac{2}{3} a \sqrt{-\frac{a}{3b}} - E_\mathrm{d}/2 \;,\nonumber
\end{eqnarray}
% TA1
 resulting in characteristic temperatures of
\begin{eqnarray}
T' &=& T_0 - ( - E_\mathrm{d}/2 ) ^ {2/3} (27 b / 4)^{1/3} / a_0 \label{eq:character} \,, \\
T'' &=& T_0 - ( E_\mathrm{init} + E_\mathrm{d}/2 ) ^ {2/3} (27 b / 4)^{1/3} / a_0. \nonumber
\end{eqnarray}

\section{Results and Discussion} \label{sec:results}
In the following, we discuss systematically how anti-parallel internal fields modify the ECE.
Hereby, we show that the ECE depends on the relative strength of the external and internal fields. 
In Subsec.~\ref{subsec:density} we realize different ratios of both fields by a variation of the bias field ($E_\mathrm{d}$) for the removal of an external field with fixed magnitude ($E_\mathrm{init}=0.2$ and $E_\mathrm{end}=0$).
Supplementarily, the variation of the external field for a fixed induced bias field ($E_\mathrm{d}=-0.6$) is discussed in Appendix~\ref{sec:fieldECE}.

%%%%%%%%%%%%%%%%%%%%%%%%%%%%%%%%% results
\subsection{Influence of internal bias fields} \label{subsec:density}
\begin{figure*}[htb]
	\centering 
 	\centerline{\includegraphics[height=0.4\textwidth]{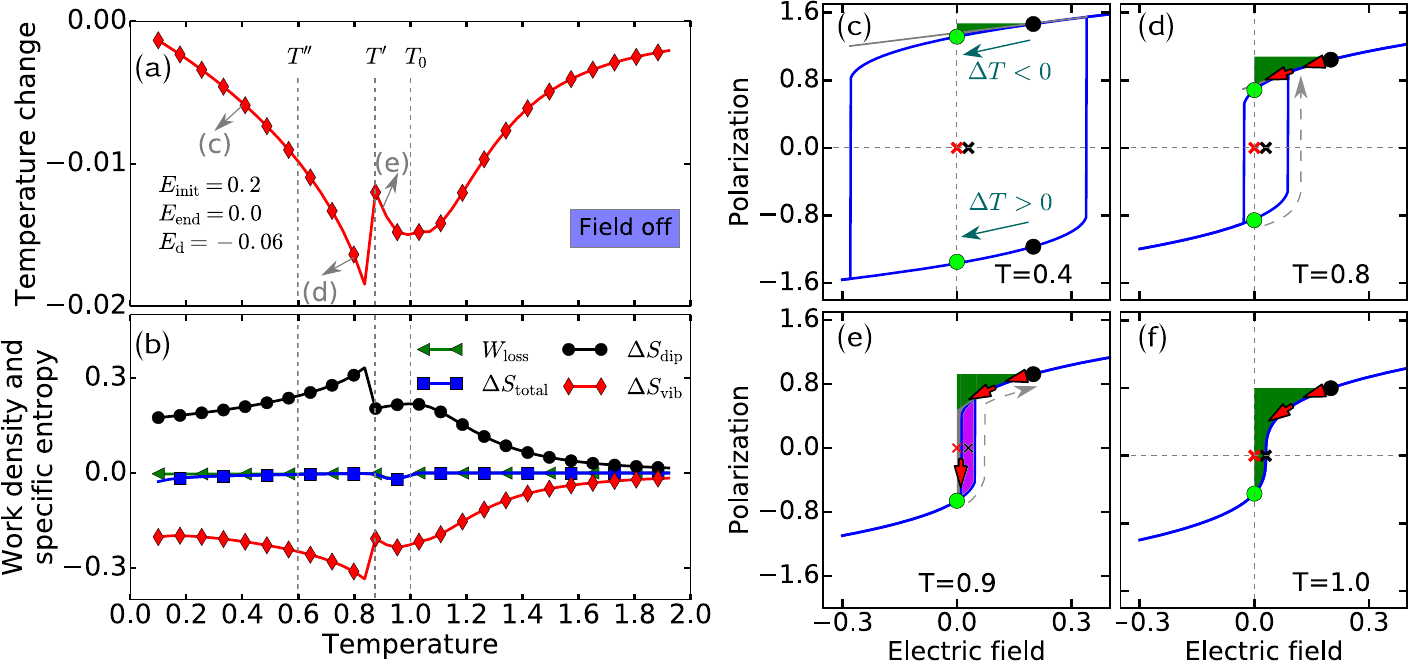}}
	\caption{ \textbf{Case (ii):} Electrocaloric effect for weak internal bias fields $ 0.06 = |E_\mathrm{d}| < E_\mathrm{init}=0.2 $ in the sample with positively poled initial states. 
		Subfigures illustrate (a) the temperature change $\Delta T$, (b) the corresponding work loss density and specific entropy changes and (c)-(f) representative hysteresis loops. In (c)-(f) dots, crosses and arrows illustrate initial and final state, the center of the hysteresis, and the direction of the field change. Ramping off the field, (c) demonstrates the conventional ECE on the upper branch, and the inverse ECE on the lower branch.
	} 
	\label{fig:defect2}
\end{figure*}

\begin{figure*} [htb]
	\centering 
	\centerline{\includegraphics[height=0.4\textwidth]{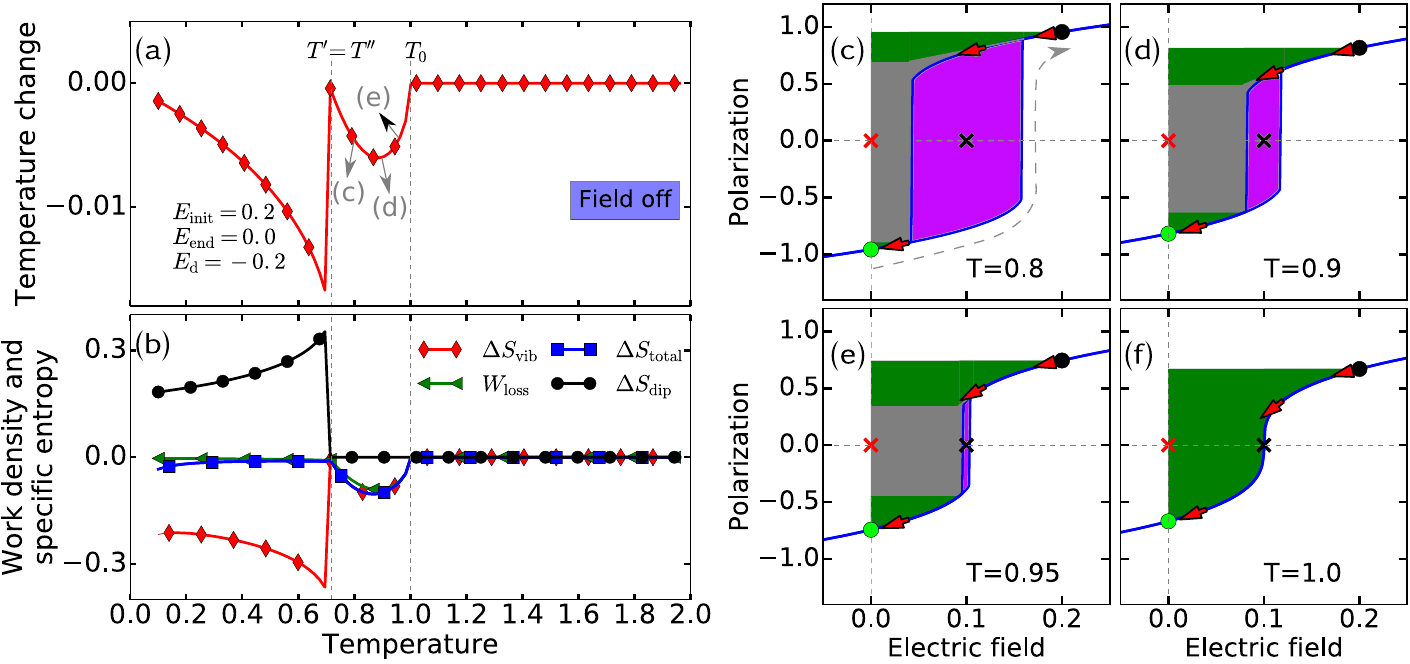}}
	\caption{
		% % % % % % general
		\textbf{Case (iii):} Electrocaloric effect for equal internal and external field strength
		($|E_\mathrm{d}| = E_\mathrm{init}=0.2$) in the sample with positively poled initial states.
%		 \AG{Maybe add for the example of a positively poled initial state?} 
		 		 Subfigures illustrate (a) the temperature change $\Delta T$, (b) the corresponding work loss density and specific entropy changes and (c)-(f) representative hysteresis loops. In (c)-(f) dots, crosses and arrows illustrate initial and final state, the center of the hysteresis, and the direction of the field change.
	}
	\label{fig:defect3}
\end{figure*}

\begin{figure*}[t]
	\centering 
	\centerline{\includegraphics[height=0.4\textwidth]{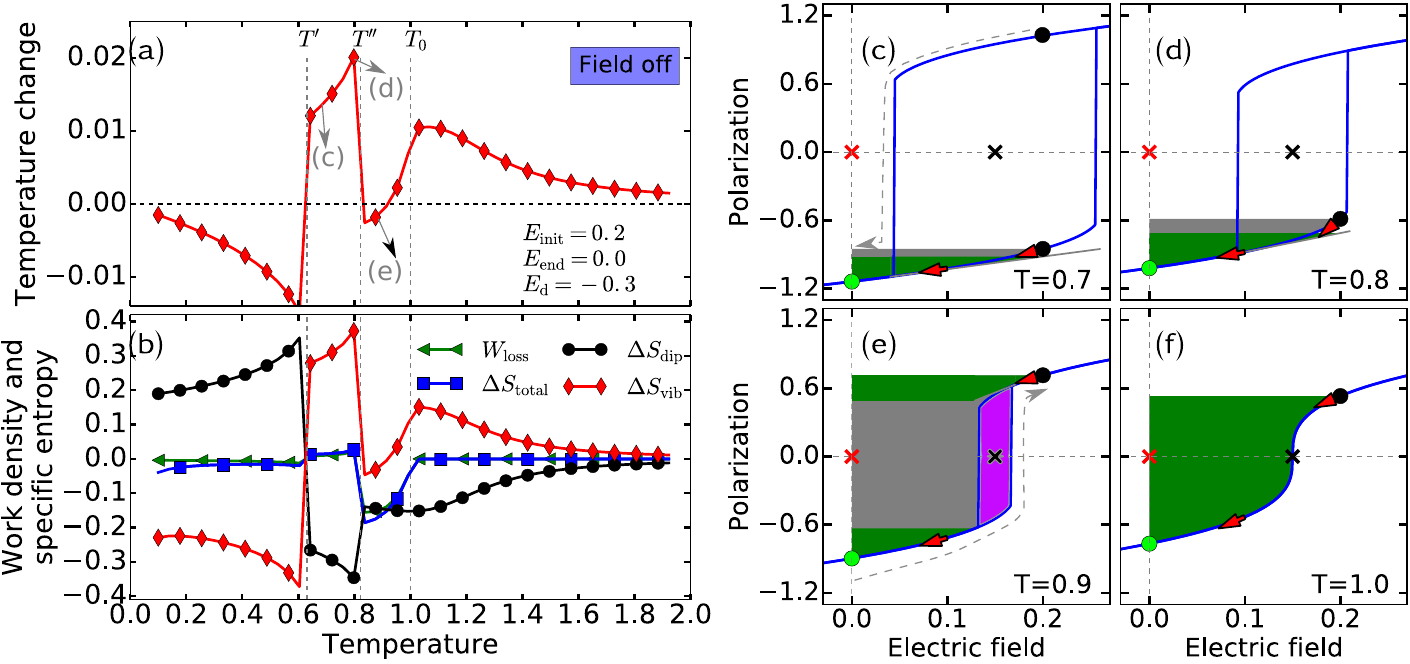}}
	\caption{
		% % % % general
		\textbf{Case (iv):} Electrocaloric effect for stronger internal bias fields ($ 0.3 =| E_\mathrm{d}| >E_\mathrm{init} = 0.2$) in the sample with negatively initial poled states for $T'<T<T''$, and positively initial poled states for all other $T$.
%		\AG{Maybe add: Exemplary, negatively poled and positively poled initial states have been chosen for $T'<T<T''$ and all other T, respectively.} 
		Subfigures illustrate (a) the temperature change $\Delta T$, (b) the corresponding work loss density and specific entropy changes (b) and (c)-(f) representative hysteresis loops. In (c)-(f) dots, crosses and arrows illustrate initial and final state, the center of the hysteresis, and the direction of the field change.
	}
	\label{fig:defect4}
\end{figure*}

\begin{figure*}[htb] 
	\centering 
	\centerline{\includegraphics[height=0.4\textwidth]{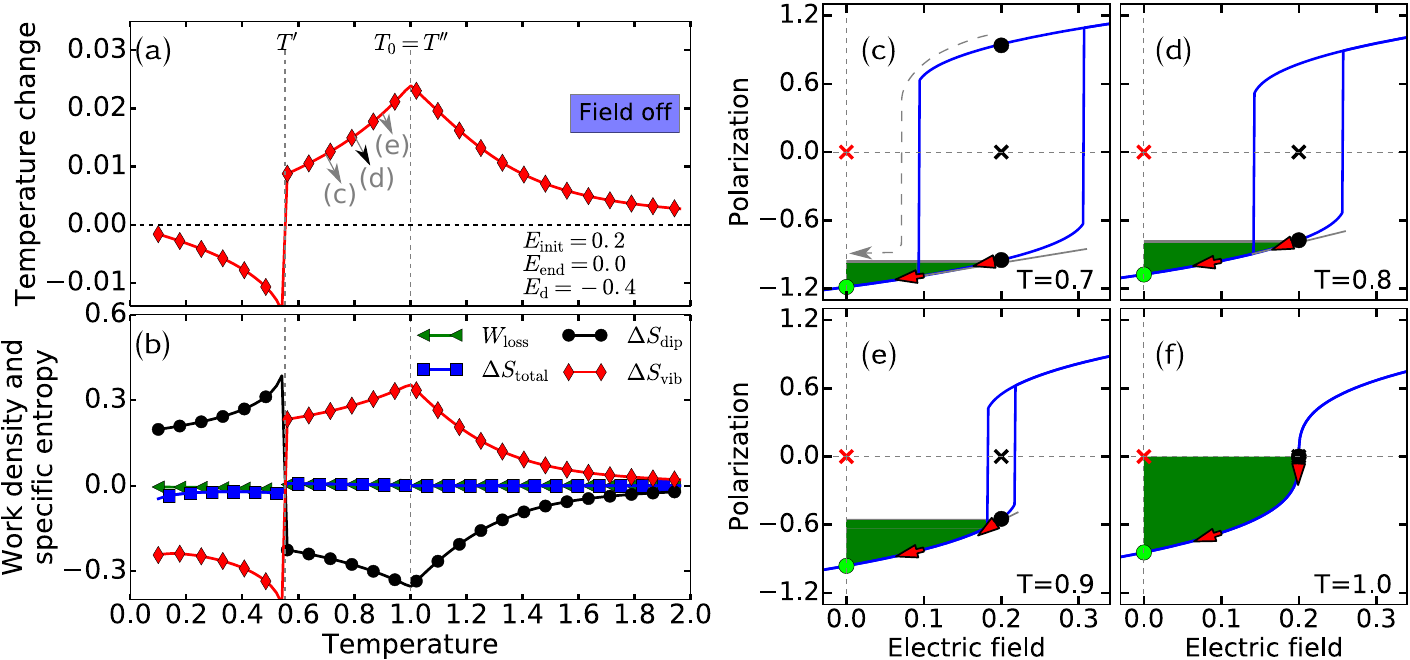}}
	\caption{% % % % general
		\textbf{Case (v):} Electrocaloric effect for strong internal bias fields ($|E_\mathrm{d}|=2E_\mathrm{init}=0.4$) in the sample with positively initial poled states for $T<T'$ and negatively initial poled state for $T>T'$.
%		 \AG{Maybe add: 
%Exemplary, negatively poled and positively poled initial states have been chosen for $T>T'$ and $T<T'$, respectively. }
Subfigures illustrate (a) the temperature change $\Delta T$, (b) the corresponding work loss density and specific entropy changes and (c)-(f) representative hysteresis loops. In (c)-(f) dots, crosses and arrows illustrate initial and final state, the center of the hysteresis, and the direction of the field change.	}
	\label{fig:defect5}
\end{figure*}

%%%%%%%%%%%%%5 cases
It is convenient to discuss the systematic trends for five representative strengths of the internal field: (i) no internal bias fields ($E_\mathrm{d}=0$), see Fig.~\ref{fig:defect1}, 
(ii) weak internal fields $|E_\mathrm{d}| < E_\mathrm{init} $, see Fig.~\ref{fig:defect2},
 (iii) equal magnitudes of the fields $|E_\mathrm{d}|= E_\mathrm{init}$, see Fig.~\ref{fig:defect3},
 (iv) stronger internal field $|E_\mathrm{d}| > E_\mathrm{init}$, see Fig.~\ref{fig:defect4}, and 
(v) very strong internal field $|E_\mathrm{d}| = 2E_\mathrm{init}$, see Fig.~\ref{fig:defect5}.
Furthermore, the systematic reduction of the field hysteresis with temperature results in three different temperature ranges:
low temperatures (no switching of the polarization direction), intermediate temperatures (switching of the polarization direction), and high temperatures (paraelectric phase without field hysteresis).

\textbf{$\bullet$ Low temperatures: All cases}\\
The general trends for positively poled samples and {$T< T'$} can be summarized as follows.
The polarization is on the upper branch of the field hysteresis with and without the positive field. Therefore, losses are negligible during field ramping, see green triangles in Figs.~\ref{fig:defect1}--\ref{fig:defect5}~(b), and the ECE is dominated by the reversible change of the specific dipolar entropy.
For field removal, $P_\mathrm{end}<P_\mathrm{init}$, i.e., the specific dipolar entropy increases, cf.\ Eq.~\eqref{eq:deltasdip} 
 and the system cools down under adiabatic conditions. Analogous field application reduces the specific dipolar entropy, and the system heats up, i.e., a reversible conventional ECE is observed. 
 %%%%%%% mail 7a
 
For negatively poled samples, the trends for $ T < T''$ in Figs.~\ref{fig:defect1}--\ref{fig:defect5} are analogous. 
In all cases, the polarization is on the lower branch of the field hysteresis with and without the positive field, losses are negligible, and the ECE is dominated by the reversible change of the specific dipolar entropy. As $|P_\mathrm{end}| > |P_\mathrm{init}|$ for field removal, the system heats up and vice versa for field application, i.e., a almost reversible inverse ECE occurs.
For field removal, Figs.~\ref{fig:defect1}(c) and ~\ref{fig:defect2}(c) demonstrate the exemplary cases with the conventional ECE on the upper branch, and the inverse ECE on the lower branch.

 With increasing strength of the internal field, $T'$ decreases from $1.0$ to $0.873$, $0.718$, $0.630$, and $0.552$ for case (i), (ii), (iii), (iv) and (v), respectively, cf. Eq.~\eqref{eq:character}.
At the same time, $T''$ increases with the ratio of internal to external field, 
 i.e., for the field strength of $E_\mathrm{init}=0.2$, we find $T''=0.552$, 0.598, 0.718, 0.822 and 1 in cases (i), (ii), (iii), (iv) and (v), respectively.
Thus, the temperature ranges with a reversible conventional and inverse ECE decrease and increase with an increasing strength of the bias field.
Well below the Curie temperature, the external field can only sample small changes of $P$, cf. Fig.~\ref{fig:defect2}~(c). In turn the ECE is small.\footnote{We note that $\Delta S_\text{total}$ scales with $1/T$. Therefore, already tiny work losses result in a finite change of the total entropy for very low temperatures.} With increasing $T$, the hysteresis narrows and the field samples larger changes of $P$. Thus both the conventional ECE for positively poled samples at $T<T'$, and the inverse ECE for negatively poled samples at $T<T''$ increase with temperature, cf.\ Figs.~\ref{fig:defect1}--\ref{fig:defect5}.

\textbf{$\bullet$ High temperatures: All cases}\\
For $T>T_0$ the system is in the paraelectric phase, and thus neither work losses nor the previous poling have an impact on the ECE. We note that our simple Landau model does not account for the shift of the ECE peak to higher temperatures with increasing field strength as commonly observed in experiments and molecular dynamics simulations.

Field application and removal always result in fully reversible changes of $P$ and a reversible ECE.
In agreement with the predictions in Ref.~\onlinecite{2015_Kutnjak}., $\Delta P$ decreases systematically with increasing temperature and the ECE is maximal at $T=T_0$, see Fig.~\ref{fig:defect1}(a).

The largest conventional ECE is observed for the ideal material ($P_\mathrm{init}>0$, $P_\mathrm{end}=0$).
An increasing internal field induces an increasing negative polarization ($P_\mathrm{end}<0$) and thus systematically lowers the conventional ECE, cf. Figs.~\ref{fig:defect1} and \ref{fig:defect2}.
For $|E_\mathrm{d}|=E_\mathrm{init}$, both the positive polarization along the external field and the negative polarization parallel to the internal field have the same magnitude ($P_\mathrm{init}=-P_\mathrm{end}$). Thus the ECE vanishes, see Fig.~\ref{fig:defect3}. 
For a further increase of $|E_\mathrm{d}|$, the negative polarization without field and thus the inverse ECE increase systematically, cf. Figs.~~\ref{fig:defect3}--~\ref{fig:defect5}.

Thus, for both low and high temperatures, the relative strengths of both fields determines the magnitude and the sign of the reversible ECE.
Anti-parallel internal fields systematically reduce the conventional ECE, whereas the opposite holds for the inverse ECE .

\textbf{$\bullet$ Intermediate temperatures}\\
For intermediate temperatures, already unipolar field cycling induces the switching of the polarization direction
from the upper to lower branch ($T>T'$) and from the lower to upper branch ($T>T''$) giving rise to a complex dependency of the ECE on relative field strengths and previous poling.
% and we discuss the different cases separately in the following.

\textbf{$\ast$ Case (i): $|E_\mathrm{d}|=0$}\\
Without internal bias fields and for the chosen strength of the external field, the characteristic temperatures order as $T''<T'=T_0$, cf. Eq.~\eqref{eq:character}. 
%Thus, no intermediate temperature range is observed for positively poled samples.
Ramping on a positive field on a negatively poled sample in the temperature range $T''<T<T'$ induces the switching of the polarization direction, cf.  the gray dashed arrow in Fig.~\ref{fig:defect1}(d). This switching reduces $\Delta |P|$ and in turn $\Delta S_\mathrm{dip}$. Furthermore positive work losses are induced.
However, this response is not accessible in a cycling field as the system would stay in the positively poled state after the first field pulse.
%We should maybe not write the following, if we want to conclude that the irreversible design option 5 can be a benefit?\\
%However, this response is most likely not useful in a cycling device as the system stays in the positively poled low temperature state after the first field pulse.\\
Hence, this response is 
not useful in a cooling device.

%For $T>T'=T_0$ the system is in the paraelectric phase.
%In this temperature regime, field application and removal always result in an increase or decrease of $P$ without switching of the polarization direction and a reversible conventional ECE.
%We note that our simple Landau model cannot account for the shift of the ECE peak to higher temperatures with increasing field strength as commonly observed in experiments and molecular dynamics simulations.
%In agreement with the predictions in Ref.~\onlinecite{2015_Kutnjak}., $\Delta P$ decreases systematically with increasing temperature and the ECE is maximal at $T'=T_0$.

\textbf{$\ast$ Case (ii): $|E_\mathrm{d}|<E_\mathrm{init}$}\\
% % % % % % % % % % % fairly low magnitude of defect
Already weak bias fields (e.g., $|E_\mathrm{d}|=0.06$) modify the characteristic temperatures ($T''<T'<T_0$), and impose strong modifications of the ECE, see Fig.~\ref{fig:defect2}.

For $T \geq T'$, the cycling of the unipolar field results in a repeatable switching between positive and negative polarization direction. As the switching reduces $\Delta |P|$ and in turn $\Delta S_\mathrm{dip}$.
$\Delta T$ is abruptly reduced at $T'$, compared to the response at lower $T$, as illustrated for the field removal.

%Within the paralectric phase ($T>T_0$) the ECE is systematically reduced with temperature and smaller compared to case~(i).
%For $T''<T<T'$, the direction of the polarization does not switch for field application and removal, resulting in a reversible change of the dipolar entropy during field cycling.

As the left coercive field is small for weak bias fields, $W_\mathrm{loss}$ and $\Delta S_\mathrm{total}$ are negligible in particular for field removal. Although, the positive work loss density increases by the area of the field hysteresis for field application, this area is small
as $T' \lesssim T_0$, see Fig.~\ref{fig:defect2}~(e). Thus no large differences in $|\Delta T|$ have to be expected between cooling and heating in a cycling field. 
%Even better: Could you maybe calculate the actual difference at least for one exemplary temperature? }

Analogous tendencies of specific dipolar entropy change and work loss density occur for field application in a negatively poled sample with $T''<T<T'$ (not shown in Fig.~\ref{fig:defect2}(a)).
Here, an increasing work loss density has to be expected with decreasing $T$ due to the increasing width of the thermal hysteresis. As discussed in case (i), the corresponding ECE is however only relevant for the first field pulse and would irreversibly heat up the material, cf. the gray dashed arrow in Fig.~\ref{fig:defect2}(d).

With increasing strength of the internal bias field, the specific dipolar entropy change for $T>T'$ is systematically reduced. Losses and their difference for field application and removal increase systematically.

%%%%%%% moderate magnitude of defect

\textbf{$\ast$ Case (iii): $|E_\mathrm{d}|= E_\mathrm{init} $}\\
When the internal bias field reaches the external field strength, $\Delta |P|$ for the field induced switching of the polarization direction is zero and $T''=T'$. Hence, $\Delta S_\mathrm{dip}$ is zero in the intermediate temperature regime and there is no difference between previous positive or negative poling. 
Nevertheless, the work losses result in a finite ECE. For field removal the losses induce the cooling of the material (conventional ECE). 
%\AG{Maybe this should be underlined a bit more as it is one of the most important new findings of our paper??? e.g.:\\
In contrast to the commonly observed heating by irreversible losses, a unique feature for $E_\mathrm{d}>E_\mathrm{init}$ is the negative work loss, i.e., the cooling of the material during field removal. This rather counterintuitive finding is related to the positive left coercive field.

The temperature dependency of the losses for field removal is illustrated by grey areas in Figs.~\ref{fig:defect3}~(c)--(e).
With increasing $T$, due to the reduction of 
thermal hysteresis and irreversible polarization changes, the width of the gray area increases, while the height decreases. 
This results in a local maximum of the losses and the corresponding ECE between $T'$ and $T_0$, see Figs.~\ref{fig:defect3}~(a).
 
%\AG{As discussed in the mail, I would not write the following and rather only say; Thus, $\Delta T$ is zero at T' and for $T\geq T_0$ and is "v"-shaped in between.} Thus, the overall $\Delta T (T)$ profile for field removal is "W-shaped".

%%%%%%% mail 7 field on and off
It should be noted that $|W_\mathrm{loss}|$ further increases for field application due to the finite width of the field hysteresis, cf. the gray dashed arrows in Fig.~\ref{fig:defect3}(c). Therefore, the heating found for field application is exceeding the cooling for field removal due to the extra work losses, cf. the pink area in Figs.~\ref{fig:defect3}~(c)--(e).
Due to the reduction of thermal hysteresis with $T$, this difference vanishes gradually. 
%\AG{again maybe one value, e.g. 
%at T=... the maximal cooling is ... whereas the heating ....
%}

% % % % % % % % % % 

\textbf{$\ast$ Case (iv): $|E_\mathrm{d}|> E_\mathrm{init} $}\\
%%%%%%% strong magnitude of defect
As soon as the internal field exceeds the external field, $T''$ is larger than $T'$, cf. Eq.~\eqref{eq:character}.
For $T>T''$, the polarization direction switches for application and removal of the positive field,	and the previous poling of the sample has no impact on the ECE.
As $|E_\mathrm{d}|> E_\mathrm{init} $, the negative polarization without external field exceeds the positive polarization induced by the external field, see Fig.~\ref{fig:defect4}~(e). Thus $\Delta S_\mathrm{dip}$ is negative (inverse ECE) and has a local maximum at $T_0$.
At the same time, a conventional ECE is induced by the losses. 
As $W_\mathrm{loss}$ is maximal at $T''$ and gradually decreases to zero at $T_0$, the net temperature change under field removal varies from a small negative value (conventional ECE) around $T''$ to a large positive value (inverse ECE) at $T_0$.

Field application induces larger losses, cf. the gray dashed arrow and the pink area in Fig.~\ref{fig:defect4}~(e), and in particular close to $T''$ an enhanced net conventional ECE would occur, heating up the sample.
% \AG{Again, maybe give numbers? Otherwise I would say would occur? Maybe we should say here that thus the sample would heat up in a cycling field?}
 With increasing $T$ the thermal hysteresis vanishes and thus also the difference between field application and removal becomes negligible.

We note that the same trends (switching of the polarization direction and superimposed inverse and conventional ECE) also occur for 
the removal of the field from a positively poled sample for $T'<T< T''$, cf. the gray dashed arrow in Fig.~\ref{fig:defect4}~(c). 
%\AG{Does this result in an overall cooling or heating? If it is overall cooling we maybe should mention it? At least this would allow for a cooling in a non-cylcing device ,too?} 
However, in this temperature range further cycling of the field cannot induce the back switching of the polarization. Hence, the reversible inverse ECE discussed for negatively poled samples occurs in successive field cycles, cf. the red arrows.

\textbf{$\ast$ Case (v): $|E_\mathrm{d}|= 2E_\mathrm{init} $ }\\
With increasing ratio of internal to external field strength, $T''$ increases systematically and finally for $|E_\mathrm{d}|=2 E$, $T''=T_0$. In this case, a reversible inverse ECE related to $S_\mathrm{dip}$ is possible for all temperatures for the negatively poled sample.
%Analogous to the case without bias fields, $\Delta S_\mathrm{dip}$ is maximal at $T_0$ resulting in an inverse $\Delta T$-peak at this temperature.
%Above $T_0$, due to the purely reversible polarization change, the EC response is fully reversible in the sense that a constant $|\Delta T|$ has to be expected for a cycling field application and removal.
\begin{figure*}[t]
	\centering 
	\centerline{\includegraphics[height=0.28\textwidth]{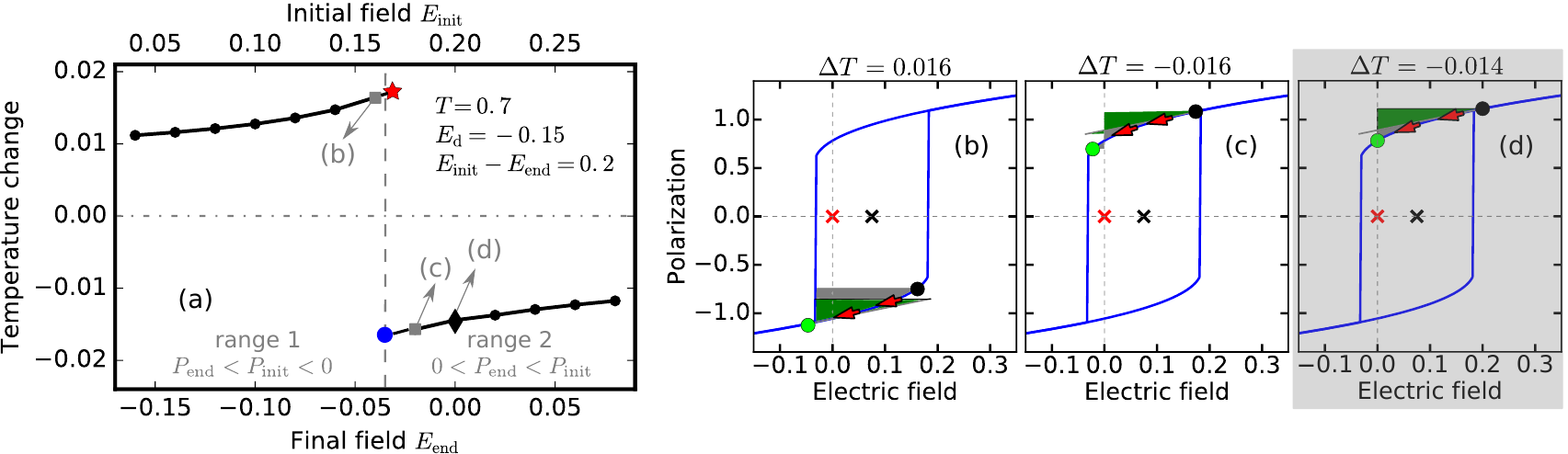}}
	\caption{Impact of field protocols on the temperature change for $T<T'$ and $E_\mathrm{init}>E_\mathrm{end}$ for case (ii): weak internal bias fields ($T=0.7$, $E_\mathrm{d}=-0.15$). 	
		(a) Temperature change for different field ranges. Representative hysteresis for the data points marked in gray are illustrated in 
		(b)--(d) and the other highlighted data points are used for Fig.~\ref{fig:pe}~(a).
		The red star and blue dot represent $\Delta T$, using {\it design option 1} and {\it 2}, respectively.
	}
	\label{fig:neg0}
\end{figure*}

As discussed for weaker bias fields, also a conventional ECE with large contributions of work losses is possible for field removal from a positively poled sample for $T>T'$, which is however only accessible in the first caloric cycle as $E<E^\mathrm{right}_c$, referring to the gray dashed arrow in Fig.~\ref{fig:defect5}(c).\\

In summary, both conventional and inverse ECE can be found in all ferroelectric materials for $T<T_0$ and
positively and negatively poled initial states, respectively.
%While the inverse ECE is commonly restricted to the ferroelectric phase and small fields in the ideal material, strong bias fields ($|E_\mathrm{init}| << |E_\mathrm{d}|$) allow for a reversible inverse ECE for all temperatures.

In the presence of bias fields, the ECE depends crucially on the relative strength of external and bias field and on temperature. First, for positively poled initial states, the bias fields reduce the temperature range and the magnitude of the reversible conventional ECE ($|E_\mathrm{init}| > |E_\mathrm{d}|$). Second, for negatively poled initial states, the bias fields increase the temperature range related to a reversible inverse ECE ($|E_\mathrm{init}| < |E_\mathrm{d}|$). 
Furthermore, $ \Delta T $ of the inverse ECE increases with an increase of the internal bias field.

For intermediate temperatures, the polarization direction switches, which reduces the dipolar entropy change and induces large losses. Excitingly, negative losses, i.e., the cooling of the material, occur for field removal and positive left coercive fields.
However, as losses and the corresponding ECE depend crucially on the direction of the field change and are commonly larger for the field application than for the field removal, the irreversible heating has to be expected in a cycling field.
% \AG{Maybe you find a more positive way to say: one has to optimize the ECE cycle if one wants to use this otherwise losses result in an irreversible heating}
The same trends are found if we change $E_\mathrm{init}$ for fixed $E_\mathrm{d}$, see Appendix~\ref{sec:fieldECE}.

\begin{figure*} [htb]
	\centering 
	\centerline{\includegraphics[height=0.28\textwidth]{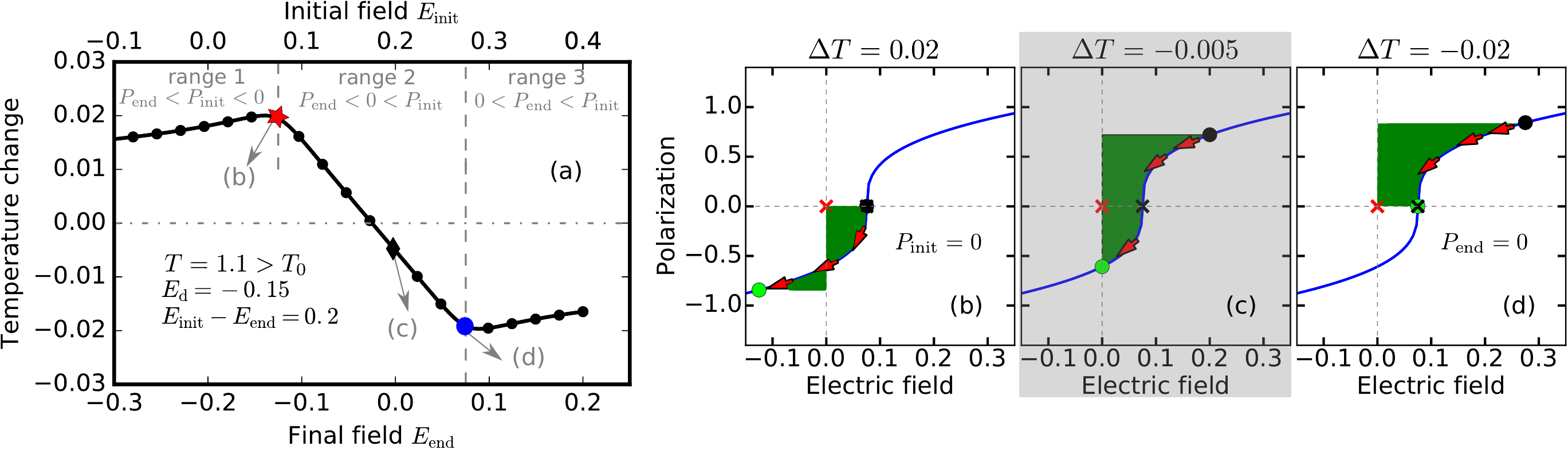}}
	\caption{Impact of field protocols on the temperature change for $T>T_0$ and $E_\mathrm{init}>E_\mathrm{end}$ for case (ii): weak internal bias fields ($T=1.1$, $E_\mathrm{d}=-0.15$). 
	(a) Temperature change for different field ranges. Representative hysteresis for the data points marked in gray are illustrated in 
		(b)--(d) and the other highlighted data points are used for Fig.~\ref{fig:pe}(a).
		The red star and blue dot represent $\Delta T$, using {\it design option 3} and {\it 4}, respectively.
	}
	\label{fig:neg2}
\end{figure*}

We note that the observed temperature- and field-strength-dependent transition between conventional and inverse ECE is in agreement with the trends which have been observed in MC and MD simulations.~\cite{2016_Mab,Grunebohm2015} 
% \AG{maybe like this?}
In particular, considering the different magnitude of the corresponding losses, i.e., the total entropy change, the results allow to interpret the larger conventional ECE found for field application than for removal.
%The seperation of 
%provide an unique view from the point of the irreversible specific entropy changes. 
%The consideration of the work losses 
% Sometimes it is instructive to separate the total en-
% tropy change into a part related to the dipolar degrees
% of freedom that give rise to the ferroelectric polarization,
% ∆Sdip , and the remaining vibrational degrees of freedom,
% ∆Slatt .9,30 Under adiabatic (and reversible) conditions,
% the change in entropy of the dipolar system (due to the
% 

%In addition, we can now interpret the large inverse ECE found for positively poled initial states and $T<T'$ for strong defect dipoles in \cite{Grunebohm2015}: The switching of the polarization direction in this case is related to a large value of $\Delta P^2$. In contrast to this, the change of $|P|$ on the lower field branch for negatively poled initial states is one order of magnitdue smaller.

%%%%%%%%%%%%%%%%%%%%%%%%%%%%%%%%% results
\subsection{Impact of the field protocol} 
\label{subsec:cycle}
\begin{figure*} [htb]
	\centering 
	\centerline{\includegraphics[width=15.5cm]{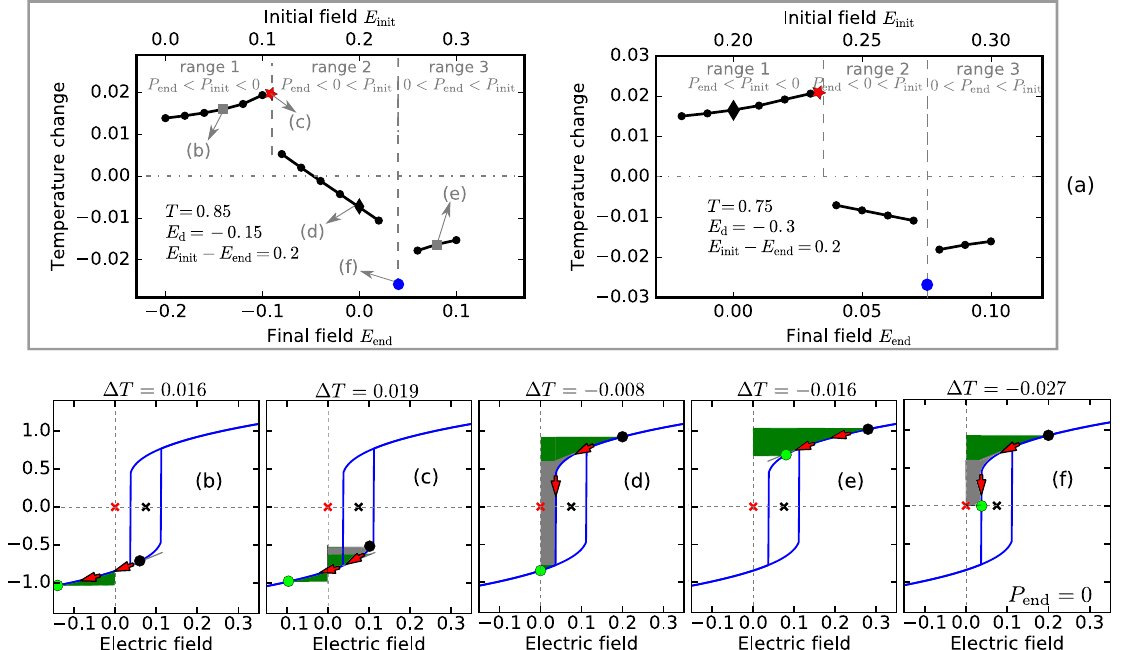}}
	\caption{Impact of field protocols on the temperature change for $T'<T<T_0$ and $E_\mathrm{init}>E_\mathrm{end}$.
		(a) Temperature change for different field ranges. Left: case (ii): weak internal bias fields ($T=0.85$, $E_\mathrm{d}=-0.15$). Right: case (iv): strong internal bias fields ($T=0.75$, $E_\mathrm{d}=-0.3$).
		Representative hysteresis for the data points marked in gray for case (ii) are illustrated in 
		(b)--(f) and the other highlighted data points are used for Fig.~\ref{fig:pe}(a).
	The red star and blue dot represent $\Delta T$, using {\it design option 2} and {\it 5}, respectively.}
	\label{fig:neg1}
	%\end{figure*}
	%\begin{figure*} [htb]
	%	\centering 
\end{figure*}

In the following section, we use the gained knowledge to tailor the ECE by the field protocol, i.e., the initial and final field strengths for a fixed field interval of $E_\mathrm{init}-E_\mathrm{end}=0.2$.
We focus on initial field strengths and polarization directions, which are accessible by a cycling electrical field of the chosen magnitude.
With $E_\mathrm{init}>E_\mathrm{end}$, for case (ii) weak internal bias fields $|E_\mathrm{d}|< |\Delta E |$ and case (iv) strong internal bias fields $|E_\mathrm{d}|> |\Delta E |$, the results are summarized in Figs.~\ref{fig:neg0}--\ref{fig:neg1}.
We note that we have chosen slightly different values of internal bias fields compared to the previous section in order to underline the general validity of our discussion.

%The dependency of the optimal ECE on $E_\mathrm{init}$ is collected for case (ii) ($|E_\mathrm{d}|=0.15<E_\mathrm{init}-E_\mathrm{end}$) and case (iv) ($|E_\mathrm{d}|=0.3>E_\mathrm{init}-E_\mathrm{end}$) in Figs.~\ref{fig:pe}.
%Figures ~\ref{fig:neg0}-\ref{fig:neg2} illustrate the underlying responses for temperature range 1 ($T<T'$), 2 ($T'<T<T_0$) and 3 ($T>T_0$), respectively. 
% general 
%For range 2 in Fig.~\ref{fig:pe}(b), in the presence of the inverse ECE, the detailed impact of field protocol is shown in Fig.~\ref{fig:neg-2}.
%%%%%%% mail 14
%\textbf{$\bullet$ Case (iv) $|E_\mathrm{d}|> E_\mathrm{init} $}\\
\textbf{$\bullet$ Low temperatures ($T<T'$)}\\%Figure~\ref{fig:neg0}~(a), Fig.~\ref{fig:neg1}~(a), and Fig.~\ref{fig:pe}(a) illustrate the impact of the field interval on the ECE for weak defects and $T<T'$, $T'<T<T_0$, and $T \geq T_0$, respectively.
At low temperatures one can depict two field intervals giving rise to repeatable and reversible responses in a cycling field.
For $E<E_\mathrm{c}^\mathrm{right}$ the system is on the lower branch of the hysteresis, and reversible 
heating is found for $E_\mathrm{init} > E_\mathrm{end}$ and cooling for $E_\mathrm{init} < E_\mathrm{end}$.
By contrast, for $E>E_\mathrm{c}^\mathrm{left}$ the system is on the upper branch, and reversible heating is found for $E_\mathrm{init} < E_\mathrm{end}$ and cooling for $E_\mathrm{init} > E_\mathrm{end}$.

Both responses can be systematically enhanced if the field samples the parts of the hysteresis with the larger slope, i.e., if
the field interval is shifted to the right for negatively poled samples, or if the field interval is shifted to the left for positively poled samples, 
as illustrated in Fig.~\ref{fig:neg0}~(c). 
Although ramping the field to the coercive fields ($P_\mathrm{end}=0$) optimizes the dipolar entropy change for both polarization directions, while at the same time positive work losses are induced, and thus an irreversible heating of the sample occurs.
These findings are in agreement to the enhancement of the ECE by reversed fields and its reduction by losses in the course of switching found for materials without bias fields,~\cite{2016_Ma,2016_Mac}, and are also verified by Molecular Dynamics simulations, cf.\ Appendix~\ref{subsec:md}.

In summary, for low temperatures ($T<T'$) and $E_\mathrm{init} > E_\mathrm{end}$, the maximal caloric heating for positively poled initial states is found for the ramping from the shoulder of the hysteresis \textbf{\emph{(design option 1)}}, cf. the red star in Fig.~\ref{fig:neg0}(a), while the maximal caloric cooling for negatively poled initial states for ramping to the shoulder of the hysteresis \textbf{\emph{(design option 2)}}, cf. the blue dot in Fig.~\ref{fig:neg0}(a).

%\textbf{$\bullet$ Design option (ii)}
%%%%%%%%%%% T>T0
%\textbf{$\bullet$ Design option (iii)}
\textbf{$\bullet$ High temperatures ($T\geq T_0$)}\\
In the paraelectric phase the response is fully reversible and three field ranges can be depicted, see Fig.~\ref{fig:neg2}.\\
In field range 1, for $E < |E_\mathrm{d}|/2$ the system is poled along the negative direction. Thus, the system heats up for 
$E_\mathrm{init} > E_\mathrm{end}$, and cools down for $E_\mathrm{init} < E_\mathrm{end}$. 
In field range 3, for $E > |E_\mathrm{d}|/2$ the system is poled along the positive direction. Thus, the system cools down for 
$E_\mathrm{init} > E_\mathrm{end}$, or heats up for $E_\mathrm{init} < E_\mathrm{end}$. 
For intermediate field strengths, the polarization direction switches for both directions of the field change.

The ECE can be maximized if the field samples the interval with the largest slope of the polarization.
Thus, the caloric heating for negative polarization can be systematically enhanced if the field interval is shifted to the right and is maximal if it approaches $|E_\mathrm{d}|/2$ \textbf{\emph{(design option 3)}}, cf.\ the red star in Fig.~\ref{fig:neg2}(a).
Analogous, the caloric cooling for positive polarization can be systematically enhanced if the field interval is shifted to the left and is maximal if it approaches $|E_\mathrm{d}|/2$ \textbf{\emph{(design option 4)}}, cf.\ the blue dot in Fig.~\ref{fig:neg2}(a).

However, in both cases if the field interval exceeds the switching field, $\Delta |P|$ and the ECE are systematically reduced. If the field interval is symmetric with respect to $|E_\mathrm{d}|/2$, i.e., for $E_\mathrm{init}+E_\mathrm{final}=E_\mathrm{d}$, the polarization switches with $\Delta |P|=0$ and there is no ECE. 
%\AG{If you like maybe comment on the dT(E) trend here?}
Thus, the external field interval has to be chosen carefully in order to obtain a maximal cooling or heating.

\textbf{$\bullet$ Intermediate temperatures ($T'<T<T_0$)}\\
The results for ferroelectric phase and intermediate temperatures are summarized in Fig.~\ref{fig:neg1}.
Analogous to the low temperature range $T<T'$, the maximal reversible heating responses for the negatively poled sample can be found if the field interval is shifted to the shoulder point of the hysteresis and $E<E_\mathrm{c}^\mathrm{right}$ \textbf{\emph{(design option 1)}}, cf. the red star in field range 1 of Fig.~\ref{fig:neg1}(a).
Analogous, the maximal cooling response for the positively poled material can be optimized if the field interval is shifted to the shoulder of the hysteresis 
and $E>E_\mathrm{c}^\mathrm{left}$ \textbf{\emph{(design option 2)}} in field range 3 of Fig.~\ref{fig:neg1}(a). For both cases reversing the direction of the field change reverses the sign of the response.
%\AG{Maybe add statement: With temperature, the optimal field interval is shifted ....?}
With increasing the initial temperature, field ranges 1 and 3 shift to the left and right, respectively.
Meanwhile, an intermediate field range 2 appears.

In this field range 2, the polarization direction switches with $P_\mathrm{end}<0<P_\mathrm{init}$ for $E_\mathrm{end}<E_\mathrm{init}$.
Different from the sole systematic change of the dipolar entropy with the field interval discussed for the paraelectric phase, the polarization direction-switching induces large work losses. Due to the large change of $P$ in the course of switching, the losses are governed by the small field interval close to $E_\mathrm{c}$.
For $T>T'$ the left coercive field is positive and thus losses further cool the material for $E_\mathrm{end}<E_\mathrm{init}$. 
%\ma{Thus, the $\Delta T(E)$-curve is systematically shifted towards negative temperatures. I delete this sentence. It is only true if we don't consider dipolar entropy change.}

%%%% special case
This opens up the possibility to combine work loss and optimal dipolar entropy change in order to obtain a maximal overall cooling \textbf{\emph{(design option 5)}}, cf. the blue dot in Figs.~\ref{fig:neg1}(a) and (f).
First, the dipolar entropy change is maximal for $P_\mathrm{end}=0$, i.e., for $E_\mathrm{end}=E_\mathrm{c}^\mathrm{left}$.
Second, the field induced change of the $P$ is related to an enhanced negative work loss.
Switching $P_\mathrm{end}$ to the negative direction, negative losses would be further enhanced, contributing more to the overall cooling.
However, at the same time the dipolar entropy change would be reduced. 
Therefore, the point of maximal cooling might appear beyond $E_\mathrm{c}^\mathrm{left}$, and depends crucially on materials systems, field strengths and temperatures.
For simplicity, this is not discussed in the current paper.

%.\AG{Add discussion here: work loss for Pinit to P=0 and work loss for the complete switching path. Which field strength would be ideal for both entropy contributions?}

This is an unique feature of systems with internal bias fields anti-parallel to the initial poling. In contrast to that, for a negative $E_\mathrm{c}^\mathrm{left}$ found without bias fields, when a negative field is applied, the work loss is positive and induces irreversible heating. In addition, also the inverse ECE found on the lower branch of the field hysteresis cannot be optimized by {\it design option 5}. In this case, the cooling found for the opposite direction of the field change would be reduced by the positive work loss.

We note that {\it design option 5} can always enhance the cooling response for positively poled initial state and $0<E_\mathrm{end}<E_\mathrm{c}^\mathrm{left}$. For $E_\mathrm{init}<E_\mathrm{c}^\mathrm{right}$, the response is however not repeatable for a cycling field of the chosen interval.
We note that a similar irreversible response may be found at first order phase transitions (thermal hysteresis).~\cite{Marathe2018}
Furthermore, as discussed in the previous section, the work loss for the reversed field direction imposes an even larger heating of the sample. 
Regarding this apparent difference of heating and cooling, considerable attention should be given in the thermodynamic design of devices, which is beyond the scope of this paper.

%It is also true in defect-free samples due to the cooling and heating difference caused by the unavoidable thermal hysteresis effect.

%the extra cooling related to the work loss is per definition not reversible. Thus, 

The same trends can be found for weak and strong bias fields, cf. left and right part of Fig.~\ref{fig:neg1}~(a). 
Thus, in both cases all three field ranges and comparable maximal cooling and heating occur. However, the different magnitude of the internal field strength induces a shift of the field intervals to positive fields.

%\AG{May one also add short comments on the change of the Wloss and dT with the temperature and internal field strength?}

\begin{figure} [htb]
	\centering 
	\centerline{\includegraphics[width=6cm]{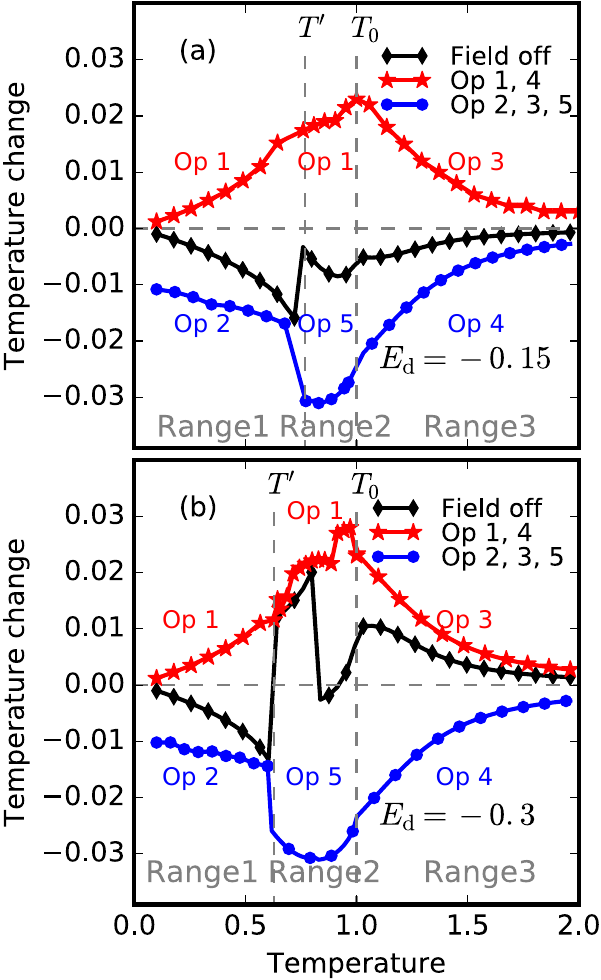}}
	\caption{ECE with $E_\mathrm{init}-E_\mathrm{end}=0.2$ for internal bias fields (a) $E_\mathrm{d}=-0.15$ and (b) $E_\mathrm{d}=-0.3$.
		%%%%%%%%%
		Black: $E_\mathrm{end}=0$, for the initial poling direction discussed in Subsec.~\ref{subsec:density}. Red: design options 1, and 4 with optimized heating response. Blue: design options 2, 3 and 5 with optimized cooling response. 	
	}
	\label{fig:pe}
\end{figure}

%% general
In summary, the reversible ECE can be found for ramping from or to zero polarization ($T>T_0$), or for the ramping from or to the shoulder of the hysteresis ($T<T_0$). Furthermore, the cooling can be enhanced by the irreversible work loss for $T'<T<T_0$, i.e., positive left coercive fields.
The optimized ECE for $ E_\mathrm{init} > E_\mathrm{final}$ is compared to the response found for simple field removal in Fig.~\ref{fig:pe}. 
For simplicity, only those responses which give rise to a repeatable cooling in a cycling field of the given field strength are included.

Obviously, the proposed design strategies allow to considerably modify the ECE. 
First of all, the reduction of the ECE by internal bias field found for simple field removal can be avoided, comparing black with red and blue lines.
Second, for small field strengths, the optimized cooling in the ferroelectric phase ($T < T_0$), i.e., in an attractive temperature range for most applications, can compete with the ECE for $T>T_0$.
In particular, the combination of reversible and irreversible contributions ({\it design option 5}) allows for a large cooling in a broad temperature range, which further broadens with increasing internal fields.
Third, even the sign of the response can be reversed by choosing proper field interval.

As discussed in Refs.~\onlinecite{2016_Mab,2012_Ponomareva}, one may enhance the ECE by the combination of inverse and conventional ECE. 
Similarly, based on the results discussed above, according to {\it design options 1} and {\it 3} the optimal cooing can be also obtained for $E_\mathrm{init}<E_\mathrm{final}$, and can be combined with the optimal cooling though {\it design options 2}, {\it 4} and {\it 5} for $E_\mathrm{init}>E_\mathrm{final}$. Additionally, one can enlarge the overall temperature span with large ECE by introducing internal bias fields.

\section{Conclusion} \label{sec:conclusion}

We delicately studied the ECE in the presence of internal bias fields and revealed the complex dependency of the response on temperature, relative field strengths and field protocol. 
In the paraelectric phase, or if the system is on one branch of the ferroelectric hysteresis without the switching of the polarization direction, the ECE is dominated by the reversible change of the dipolar entropy. 
In case of anti-parallel internal bias fields and external fields, the relative strength of both fields determines the magnitude and the sign of the ECE. 
%Conventional and inverse ECE refers to heating/cooling by increasing the electric field, or cooling/heating by decreasing the field.
The sample heats up in a conventional ECE by increasing the electric field, and cools down by decreasing the field.
By contrast, the sample cools down in an inverse ECE by increasing the electric field, and heats up by decreasing the field.
Already weak internal bias fields may considerably reduce the conventional ECE.
If the internal field exceeds the external field strength, a large inverse ECE can be induced.
In case of parallel internal fields, the conventional ECE is most relevant (see Appendix.~\ref{sec:para}). Thus, the combination of conventional and inverse ECE allows to enhance the overall ECE by field reversal. In case of unipolar fields, this is beneficial compared to the simple increase of the field magnitude as one can avoid ferroelectric break down and reduce Joule heating.

In case of ferroelectric switching, the work loss commonly induces irreversible heating of the material. Excitingly, a negative work loss allows to enhance the overall cooling in the presence of internal bias fields. 
Although this additional cooling is per definition irreversible but still repeatable, it may have the potential to enhance the overall cooling in a optimized thermodynamic cycle.

As summarized in Tab.~\ref{tab:opt}, we have proposed different strategies to enhance the overall responses for different temperatures by means of the field protocol. First of all, one can switch between conventional (cooling for $E_\mathrm{init} < E_\mathrm{end}$) and inverse (cooling for $E_\mathrm{init} > E_\mathrm{end} $) ECE by a shift of the field interval.
Furthermore, one can enhance the overall reversible responses, if the field interval is adjusted to the part of the hysteresis showing the largest slope. 
It is important to realize, that the temperature dependency of the thermal hysteresis results in a large temperature dependency of the optimal field intervals and simple unipolar field cycling may result in unindented switching of the polarization direction, degeneration of the ECE, or even the reversal of the induced temperature change. 
\begin{table}[t]
	\centering 
	\caption{Optimization strategies of the ECE in the presence of internal bias fields for a fixed field interval $|E_\mathrm{init} - E_\mathrm{end}|$.
	}
	\begin{tabular}{cccccc}
		\hline
		$T$&option& ECE &field interval&branch& reversible?\\
		\hline
		\multirow{2}{*}{$T<T_0$}&1 &inverse&$E<E_\mathrm{c}^\mathrm{right}$&lower&yes\\
		&2& conventional& $E>E_\mathrm{c}^\mathrm{left}$&upper& yes\\
		$T'<T<T_0$&5&conventional& $E\leq E_c$&switching&no\\
		\multirow{2}{*}{$T\geq T_0$}&3&inverse&$E\leq E_\text{d}/2$&lower&yes\\
		&4&conventional& $E \geq E_\text{d}/2$&upper&yes\\
		\hline
	\end{tabular}
	%	\centerline{\includegraphics[width=8.6cm]{11-b1-crop.pdf}}
	\label{tab:opt}
\end{table}
\begin{figure*}[ht] 
	\centering 
	\centerline{\includegraphics[height=0.4\textwidth]{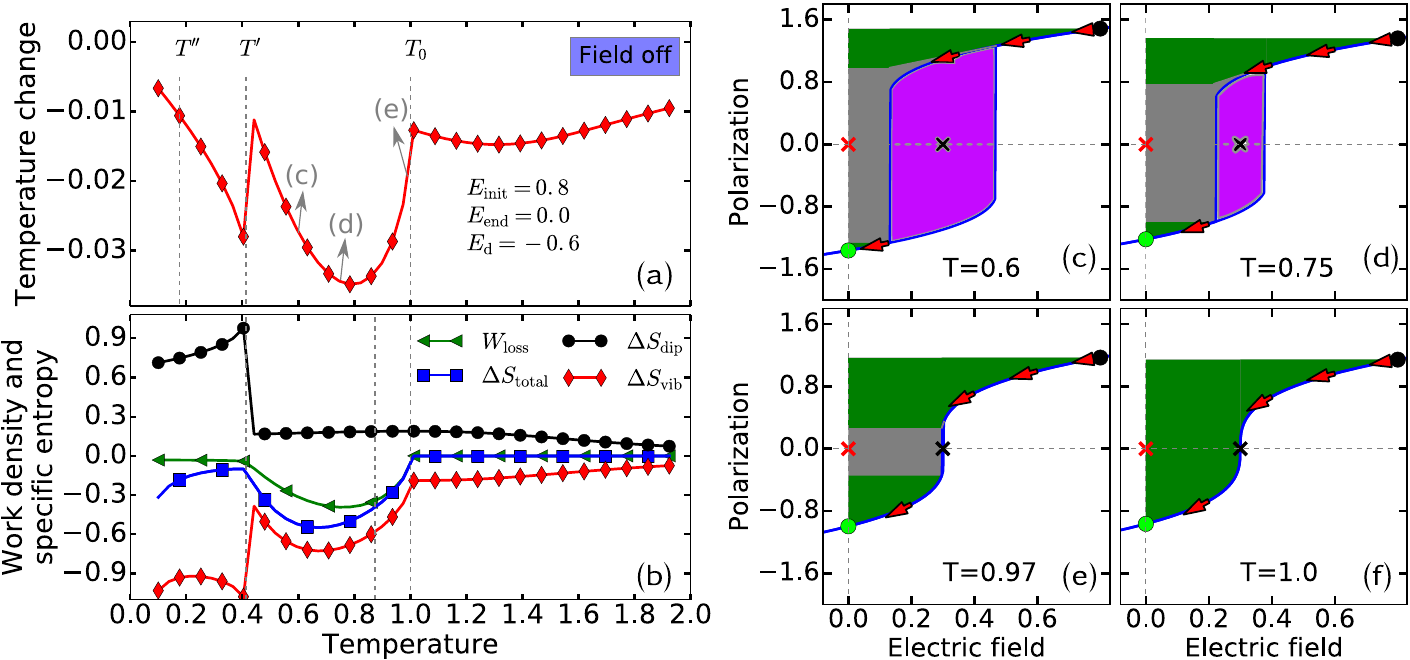}}
	\caption{Electrocaloric effect for $|E_\mathrm{d}| < E_\mathrm{init}$ in the sample with positively poled initial states, similar to case (ii) as in Fig.~\ref{fig:defect2} 
%		\AG{Maybe add for the example of a positively poled initial state?} . 
		Black and lime dots and black and red crosses mark initial and final state and the center of hysteresis without and with defects, respectively, and arrows illustrate the direction of the field change.	}
	\label{fig:field1}
\end{figure*}

We note that our analytical model does not depend on the specific origins of the internal bias fields and is thus probably applicable to a broad class of systems such as ferroelectrics with aligned defect dipoles or films and composites with imprint fields.
As proof of concept, we have compared our findings to the results based on phenomenological models and {\it ab initio}-based simulations 
for the example of aligned defect dipoles in BaTiO$_3$. Indeed our simple model can reproduce the qualitative trends and allows for a fundamental understanding of the impact of the thermal history on the ECE~\cite{Grunebohm2015,2016_Mab}.
In future, more specific models,\footnote{We note that, however, even the bias fields are not stable over long cycling periods. This may be of high technical impacts on the devices.} have to be established in order to consider the dependency of the internal fields on temperature and possible modifications with time.~\cite{2013_Erhart}
%An experimental verification of our prediction would be beneficial to perform the future research. 
Our results now have to be confirmed by experiments.
%\ma{Furthermore, an experimental verification is required to evaluate this technical impact.}

%\AG{maybe + experimental verifications???}

\begin{acknowledgments}
	The funding from Deutsche Forschungsgemeinschaft (DFG) SPP 1599 B3 (XU 121/1-2, AL 578/16-2) and A11/B2 (GR 4792/1-2) is acknowledged. 
	Additionally, the Lichtenberg-High Performance Computer at TU Darmstadt and the Center for Computational Science and Simulation (CCSS) at University of Duisburg-Essen are appreciated for the computational resources.
  We thank Dr. C. Ederer at ETH and Dr. M. Marathe at Institut de Ci\`encia de Materials de Barcelona for fruitful discussions. 
\end{acknowledgments}

\appendix
\begin{figure*}
	\centering 
	\centerline{\includegraphics[height=0.4\textwidth]{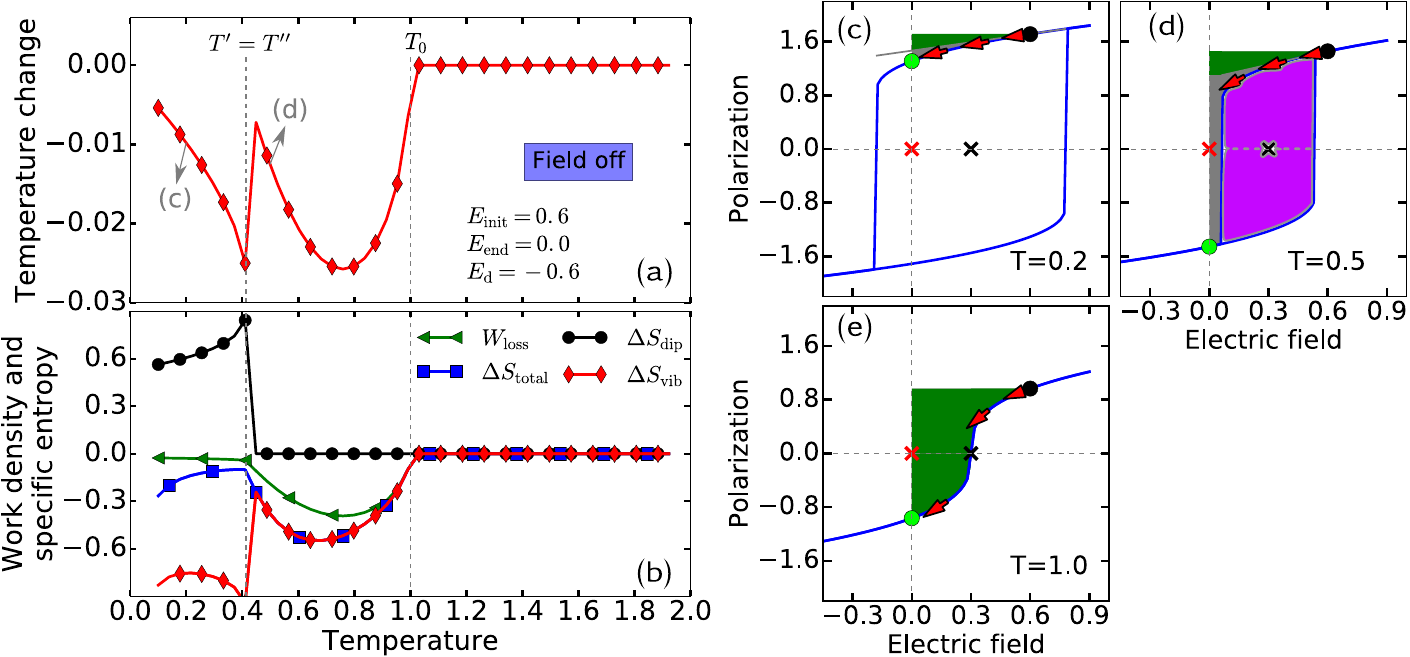}}
	\caption{Electrocaloric effect for $|E_\mathrm{d}| =E_\mathrm{init}$ in the sample with positively poled initial states, similar to case (iii).		 
%		\AG{Maybe add for the example of a positively poled initial state?} . 	
		Black and lime dots represents the initial and final polarization states, and the arrows describe the field-change directions. 
The red and black crosses stand for the center of the hysteresis without and with defects, respectively. }
	\label{fig:field2}
\end{figure*}
\begin{figure*}[t]
	\centering 
	\centerline{\includegraphics[height=0.4\textwidth]{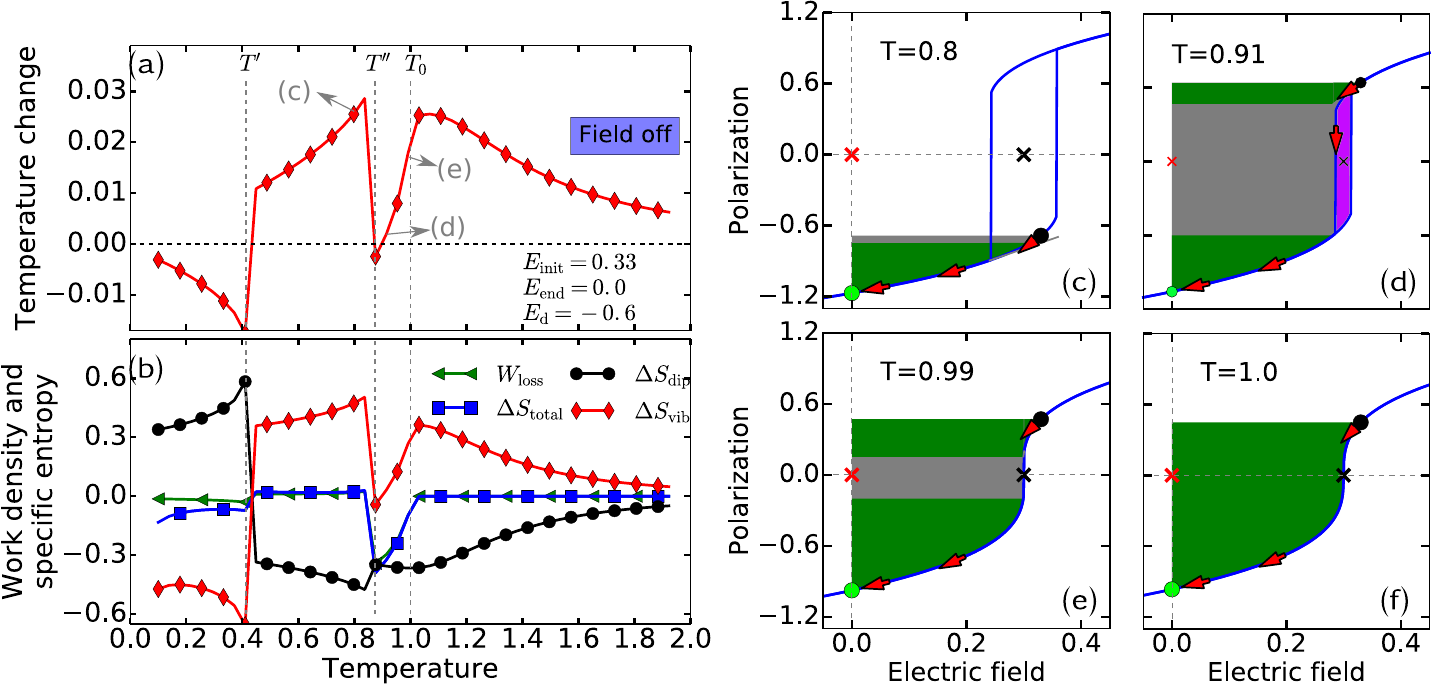}}
	\caption{Electrocaloric effect for $|E_\mathrm{d}| > E_\mathrm{init}$ in the sample with negatively initial poled states for $T'<T<T''$, and positively initial poled states for all other $T$, similar to case (iv) as in Fig.~\ref{fig:defect3}. 
%		\AG{Maybe add: Exemplary, negatively poled and positively poled initial states have been chosen for $T'<T<T''$ and all other T, respectively.} 
		Black and lime dots represents the initial and final polarization states, and the arrows describe the field-change directions. 
		The red and black crosses stand for the center of the hysteresis without and with defects, respectively. 
	}
	\label{fig:field3}
\end{figure*}
\begin{figure*}[t]
	\centering 
	\centerline{\includegraphics[height=0.4\textwidth]{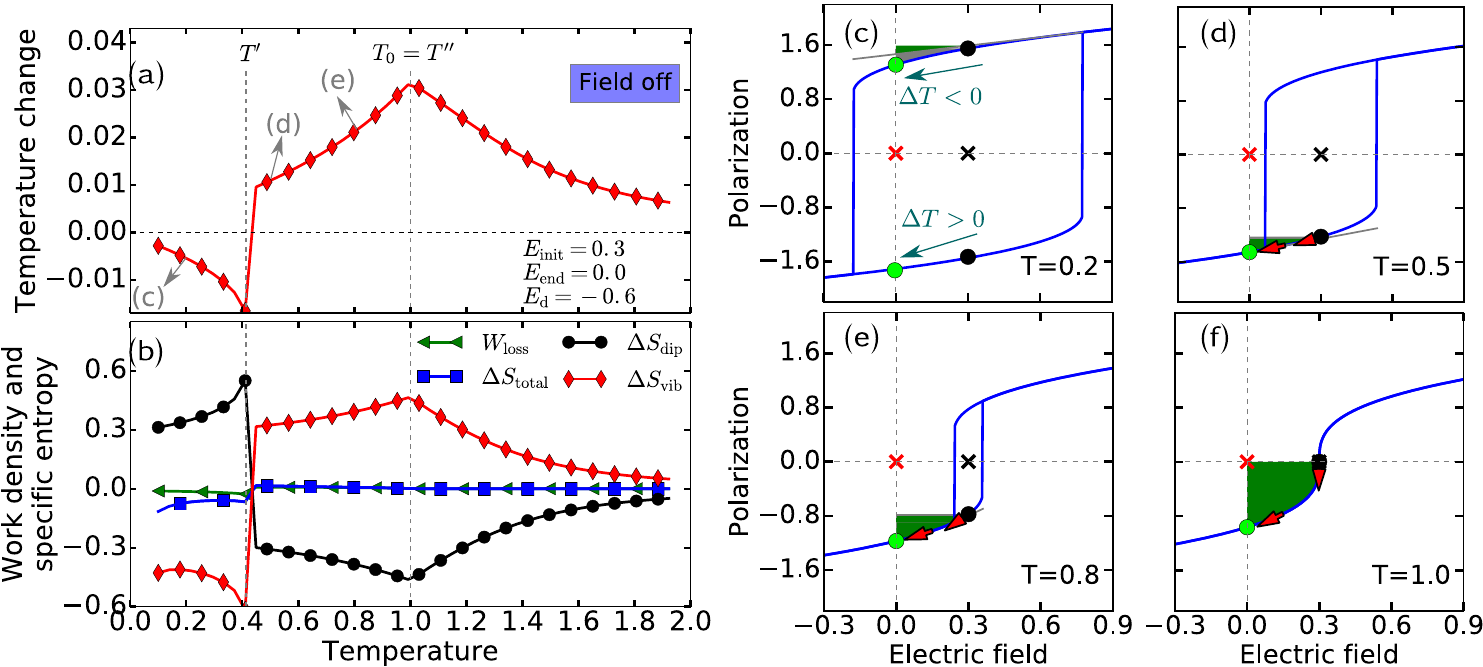}}
	\caption{Electrocaloric effect for $|E_\mathrm{d}| = 2 E_\mathrm{init}$ in the sample with negatively initial poled states, similar to case (v) as in Fig.~\ref{fig:defect4}. 
%		\AG{Maybe add: Exemplary, negatively poled and positively poled initial states have been chosen for $T>T'$ and $T<T'$, respectively.}
 Black and lime dots and black and red crosses mark initial and final state and the center of hysteresis without and with defects, respectively, and arrows illustrate the direction of the field change.
	}
	\label{fig:field4}
\end{figure*}
\section{Influence of the initial field strength} \label{sec:fieldECE}
The different trends discussed for the ECE depend on the relative strength of $E_\mathrm{d}$ and $\Delta E$.
In Sec.~\ref{subsec:density} we have studied different ratios by a systematic variation of the internal field strength for $E_\text{init}=0.2$ and $E_\text{final}=0$. In the following context we show that the same trends occur for a fixed internal field strength ($E_\mathrm{d} = -0.6$) and a variation of $E_\text{init}$. 

The results for case (ii) $|E_\mathrm{d}| < E_\mathrm{init}$ are illustrated in Fig.~\ref{fig:field1}. Analogous to Fig.~\ref{fig:defect2}, we find that the conventional ECE for positively poled samples which is reduced at $T'$ and $T\geq T_0$.
It should be noted that we use slightly different ratios of both fields (0.6/0.8 here compared to 0.06/0.2 in Sec.~\ref{subsec:density}). 
As already discussed in Sec.~\ref{subsec:density}, the reduction of $\Delta T$ is more pronounced for a reduced ratio of both field strengths.

The results for case (iii) $|E_\mathrm{d}| = E_\mathrm{init}$ are summarized in Fig.~\ref{fig:field2}. Analogous to Fig.~~\ref{fig:defect3}, the dipolar entropy change is zero for $T>T'$ as field removal results in the switching of the polarization direction with constant $|P|$.
For $E=0.6$, both the irreversible losses and the dipolar entropy change for $	T<T'$ are reduced compared to $E=0.8$.

The results for case (iv) $ |E_\mathrm{d}| > E_\mathrm{init}$ are collected in Fig.~\ref{fig:field3}. The same temperature profile of the ECE as in Fig.~\ref{fig:defect4} is observed: inverse ECE due to the dipolar entropy for $T'<T<T''$ and $T>T_0$, conventional ECE due to the dipolar entropy for $T<T'$ and the superposition of conventional (due to losses) and inverse (dipolar entropy) ECE for $T''<T<T_0$.

The results for case (v) $ |E_\mathrm{d}| = 2 E_\mathrm{init}$ are summarized in Fig.~\ref{fig:field4}. 
Analogous to Fig.~\ref{fig:defect5}, a reversible inverse ECE related to the dipolar entropy change can be found for all temperatures. Furthermore, a reversible conventional ECE can be found for $T<T'$.

%\AG{Maybe we could add and compare the different values for T' and T''?}
We note that a higher internal field strength (here $E_\mathrm{d}=-0.6$) results in an increase of the temperature range $T'<T<T_0$, compared with the discussion in Subsec.~\ref{subsec:density}.
In addition, we find $T''=$\,\,0.175, 0.413, 0.873 and 1.0 for $E_\mathrm{init}=$\,\,0.8, 0.6, 1/3 and 0.3.
In other words, $T''$ increases with decreasing the ratio $|E_\mathrm{init}/E_\mathrm{d}|$.
%, i.e., $T''$ is reduced for $Ed<Einit$

%$T'=0.413$ for the fixed internal field strength $E_\mathrm{init}=0.6$ results in an increase of the temperature range $T>T'$ in comparison to the discussion in Sec. .... 

%In addition, we find $T''=$ 0.175, 0.413, 0.873 and 1.0 for 
%$E_\mathrm{init}=$0.8, 0.6, 1/3 and 0.3, i.e., $T''$ is reduced for $Ed<Einit$....

In summary, the same tendencies are observed for the similar ratios of $E_\mathrm{d}$ and $E_\mathrm{init}$, independent on the actual values of both fields.

\section{Parallel internal and external fields } \label{sec:para}
The main focus of Sec.~\ref{sec:results} is the ECE in the presence of anti-parallel internal and external fields. However, as discussed in Subsec.~\ref{subsec:density}, shifting the field interval may considerably modify the response. In the following, we finally discuss the impact of parallel internal and external fields for $E_\mathrm{init}-E_\mathrm{end}=0.2$.
%\AG{true?} 
%
%and $E_\mathrm{end}=0$. 
Thereby, we focus on two cases: $E_\mathrm{d}=0.15<E_\mathrm{init}$ and $E_\mathrm{d}=0.3>E_\mathrm{init}$. 
\begin{figure} [tb]
	\centering 
	\centerline{\includegraphics[width=0.3\textwidth]{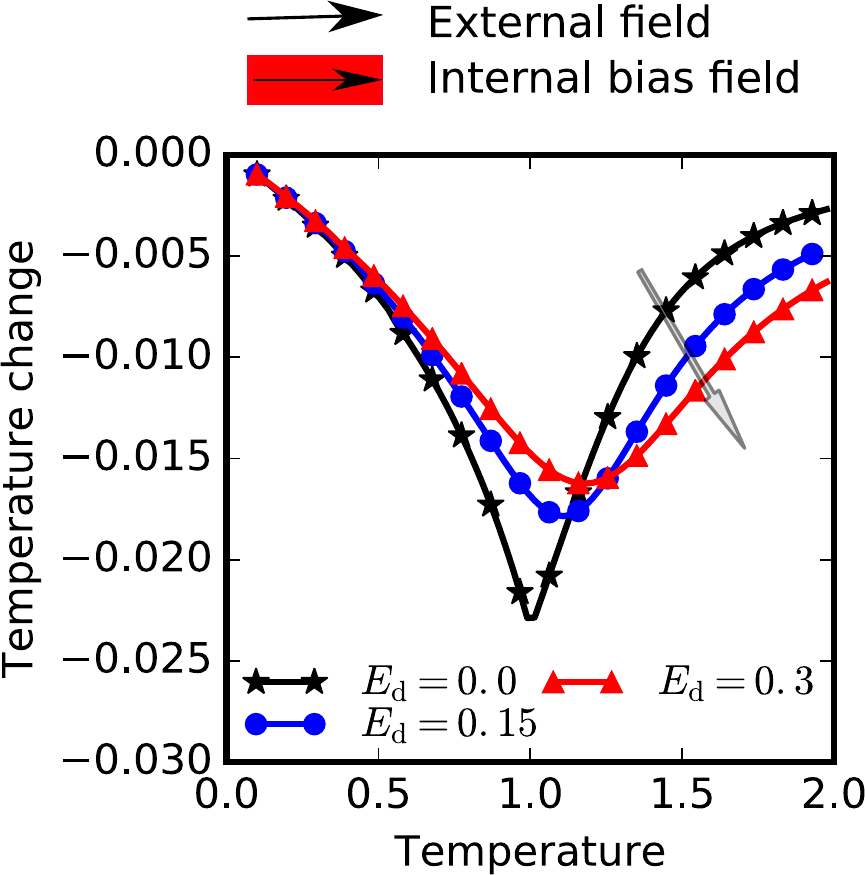}}
	\caption{Influence of parallel internal and external on the ECE for $E_\text{init}=0.2$ and $E_\text{final}=0$ in a positively poled sample. }
	\label{fig:para}
\end{figure}
In the presence of positive bias fields, the left coercive field is always negative and thus no characteristic temperature $T'$ can be defined.
In addition, $T''$ is systematically reduced from 0.55 (no internal fields) to 0.45 and 0.35 for $E_\textrm{d}=0.15$ and 0.3, respectively. 
In principle, the reversible inverse ECE for a negatively poled sample appears also on the lower branch of hysteresis, which will not be discussed in the following.

%
%the repeatable switching between both polarization direction discussed for negative internal fields for a positive unipolar field is only possible for very low temperatures and thus will not be discussed in the following.

For positively poled samples, we find a reversible conventional ECE, i.e., almost equal cooling for field removal and heating for field application. 
As shown in Fig.~\ref{fig:para}, with increasing strength of internal fields the $\Delta T$-peak position shifts to higher temperatures and broadens. Thus, $\Delta T$ above $T_\mathrm{0}$ can be enhanced by the parallel internal fields whereas the ECE for lower temperatures is slightly reduced.

These findings are in qualitative agreement to the results for parallel internal fields induced by aligned defect dipoles in Refs.~\onlinecite{Grunebohm2015,2016_Mab}. There, it has been shown that internal bias fields stabilize the ferroelectric phase and systematically shift the $\Delta T$-peak to higher $T$ with an increasing internal field strength. Furthermore, it has been discussed that the internal fields may exceed the critical field strength of the first order paraelectric to ferroelectric transition, resulting in a continuous transition.
\begin{figure*} [b]
	\centering 
	\centerline{\includegraphics[width=0.6\textwidth]{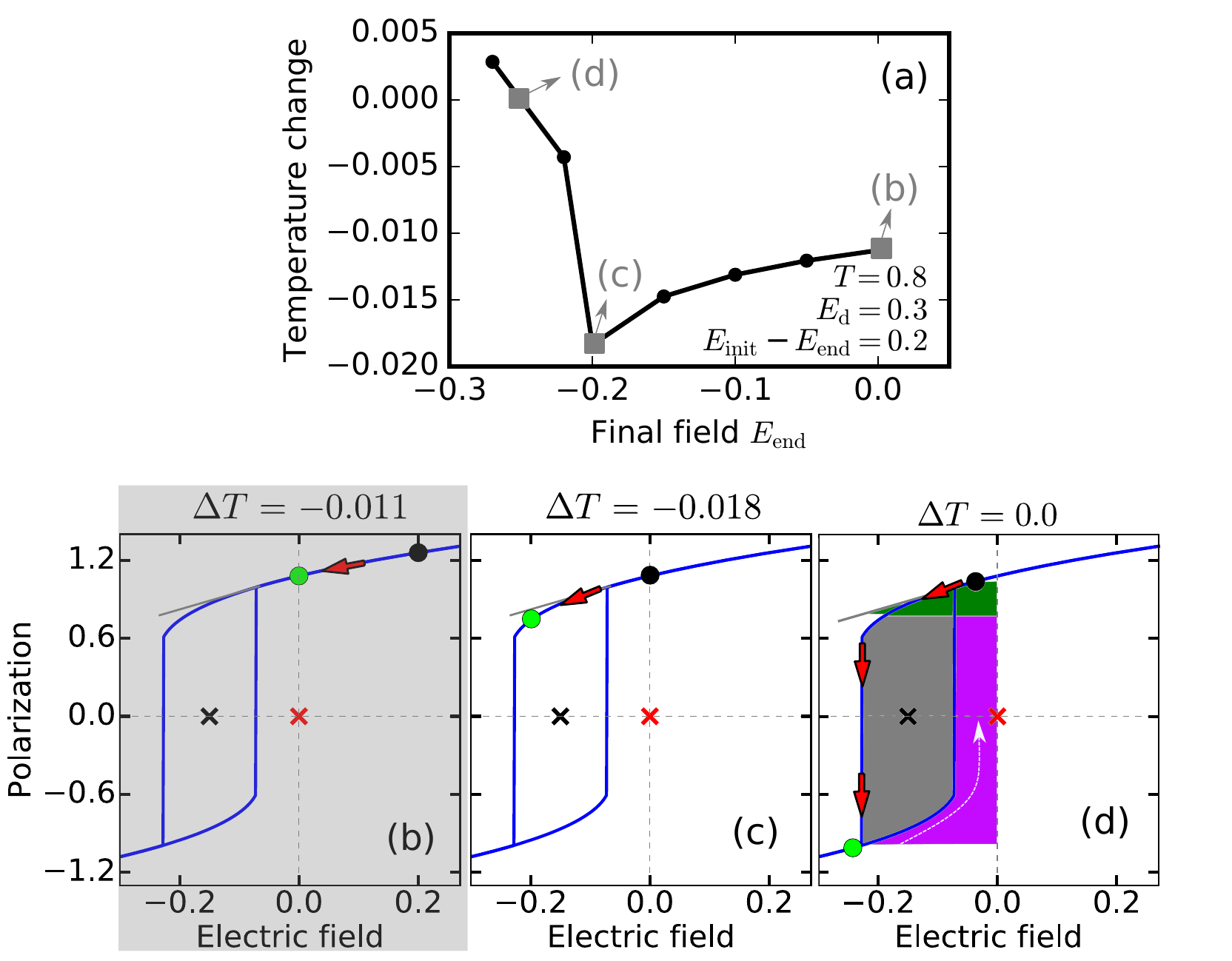}}
	\caption{Tailoring strategies of the ECE through modifying the field protocol. 		Black and lime dots represents the initial and final polarization states, and the arrows describe the field-change directions. 
		The red and black crosses stand for the center of the hysteresis without and with defects, respectively.  
	}
	\label{fig:para2}
\end{figure*}
\begin{figure*}[t]
	\centering 
	\centerline{\includegraphics[width=12cm]{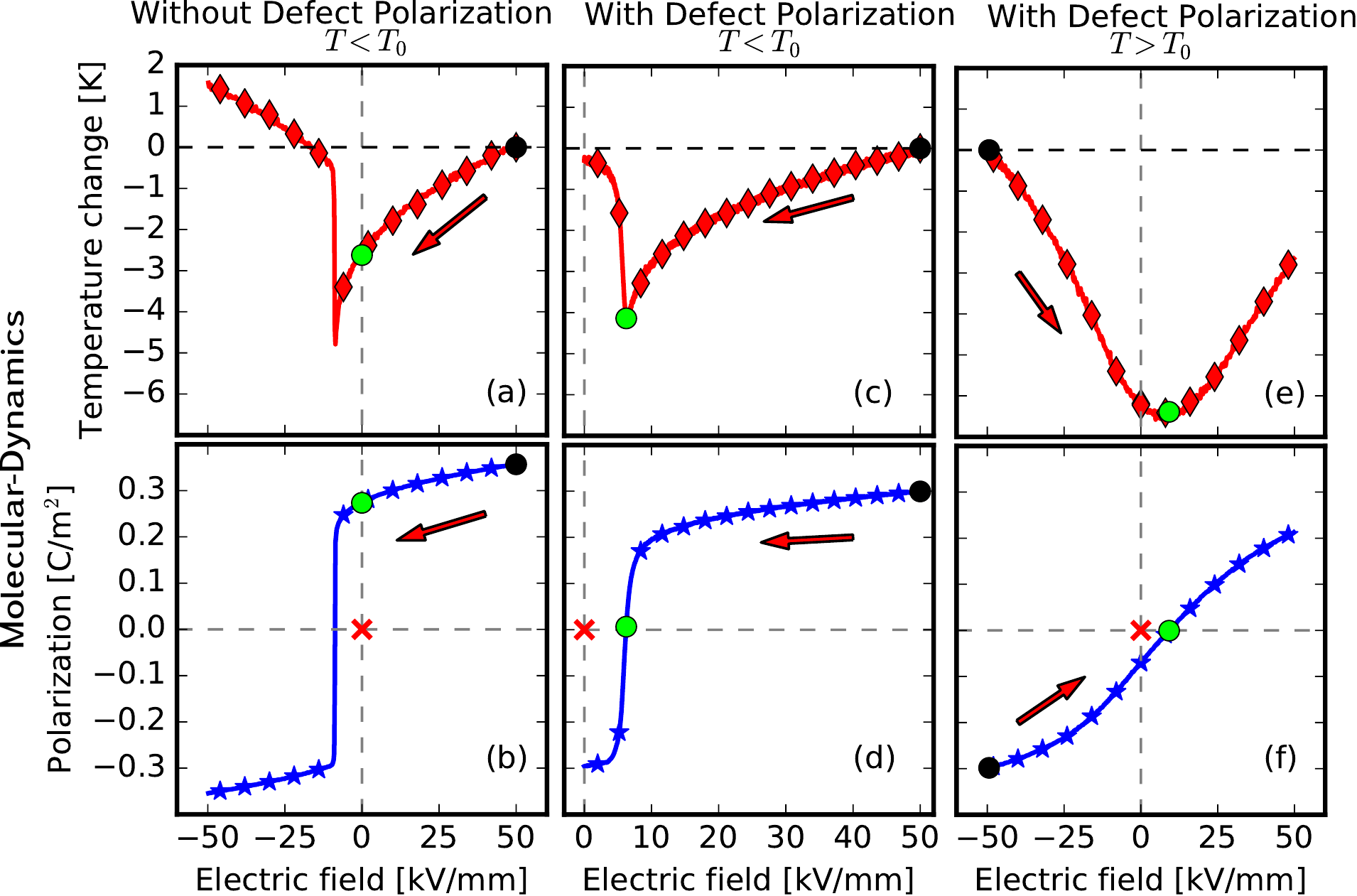}}
	\caption{
		Change of temperature (a), (c), (e) and polarization (b), (d), (f) as found in {\it ab initio}-based molecular dynamics simulations.
%		In (a)-(d) an external field is removed from a positively poled state ($E_\mathrm{init}>E_\mathrm{final}$) without and with anti-parallel internal bias fields. 
		(a)-(b) below $T_0$ without defects; (c)-(d) below $T_0$ with anti-parallel defects; (e)-(f) above $T_0$ with anti-parallel defects. The black dots represent the initial field, and the red arrows indicates direction of the loading. The center for defect-free samples are marked as the red crosses. 
	}
	\label{fig:md2}
\end{figure*}
Analogous to the discussion in Subsec.~\ref{subsec:cycle}, the reversible conventional caloric cooling can be optimized by \emph{design option 2}, i.e., if the the field interval is shifted to the shoulder of the left hysteresis, cf.\ Figs.~\ref{fig:para2}~(b) and (c).
Here, it should be noted that this shoulder point depends on temperature and the internal field strength.
With increasing strength of the internal field, the optimal field interval is thus systematically shifted to negative fields.

Similar to the discussion for negative internal fields, shifting the field interval to the left coercive field for a positively poled sample would further enhance the dipolar entropy change. However, as illustrated in Fig.~\ref{fig:para2}~(d), the switching of the polarization direction is related to large work losses and thus results in an irreversible heating of the material for the negative left coercive field. Thus, \emph{design option 5} is not applicable for positively poled samples and parallel internal fields. 
Furthermore, an optimized inverse ECE would be possible for the field interval left to the right coercive field ($E<E_\mathrm{c}^\mathrm{right}$), indicated by the white arrow in Subfig.(d), using strategy similar to \emph{design option 5}.

%\AG{Add a statement on the ECE for field application, i.e., the pink area in 17?, }

In summary, parallel internal and external fields favor the conventional ECE and the same optimization strategies as discussed for negative internal fields can be applied.

\section{Internal field related to defect dipoles -- comparison to molecular dynamics simulations}
\label{subsec:md}

In the following we compare the results based on our simple analytical model with Molecular Dynamics simulations based on \emph{ab initio} derived potentials \cite{2010_Nishimatsu} using the {\sc{feram}} code.\cite{Feram1}
We focus on internal bias fields induced by randomly distributed and perfectly anti-parallel aligned defect dipoles in BaTiO$_3$.
Details on method and technical aspects can be found in Refs.~\onlinecite{2010_Nishimatsu,2016_Marathe,2016_Grunebohm}.
In agreement to our analytical model, the center of the ferroelectric hysteresis is systematically shifted to positive fields with an increasing strength of the internal field (given by the density or strength of the anti-parallel defect dipoles). Furthermore, the thermal hysteresis is systematically reduced with increasing temperature and vanishes in the paraelectric phase.

Induced changes of ferroelectric polarization and resulting caloric responses for $E_\mathrm{init}>E_\mathrm{final}$ and a positively poled initial state are illustrate in 
Fig.~\ref{fig:md2} for three examples: (1) $T<T_0$ in defect free material in Subfigs.(a) and (b), (2) $T'<T<T_0$ in the presence of anti-parallel internal bias fields in Subfigs.(c) and (d), and (3) $T>T_0$ in the presence of anti-parallel internal bias fields in Subfigs.(e) and (f).
As discussed in Sec.~\ref{subsec:cycle}, the caloric cooling can be optimized by
\emph{design option 2} when ramping to the shoulder of the hysteresis (see the defect-free sample in Subfigs.(a) and (b)), and \emph{design option 4} ($T>T_0$) through ramping to the left coercive field (see the samples in the presence of defect dipoles in Subfigs.(e) and (f)). 

%If the field is ramped down from the positively poled sample, the polarization is reduced and in turn the material cools down due to the inverse ECE, which predicted by our analytical model.

For $T<T_0$ and a negative $E_\mathrm{c}^\mathrm{left}$, see Subfigs.(a)--(b), further shifting the field interval to the left beyond the shoulder of the hysteresis suppresses the positive polarization, and the system heats up due to the work losses, as discussed in Sec.~\ref{subsec:cycle}.
Under a further increase of the negative field, the (negative) polarization increases, giving rise to the work losses, and suppressing the available change of the specific dipolar entropy. Hence, the material heats up further.
%These results also agree with experimental observations and analytical considerations in Ref.~\onlinecite{2016_Mab,2016_Mac}.
%

Figs.~\ref{fig:md2}~(c) and (d) summarize the results for defects anti-parallel to the positively poled polarization and $T'<T<T_0$. In this case the switching of the polarization direction from positive to negative takes place at $
E_\mathrm{c}^\mathrm{left}>0$. The reduction of $\Delta P$ in combination with the irreversible negative work loss results in the maximal cooling point only slightly above the zero polarization point (\emph{design option 5}).
This agrees with the observations in Fig.~\ref{fig:neg1}(f). 
In short, a complex temperature dependency of the inverse ECE on previous poling and relative strength of internal and external field strength.\cite{Grunebohm2015,2016_Mab}
We note that the simulations furthermore also yield a reversible inverse ECE for strong internal field and $T>T_0$.

In summary, our simple analytical model allows to reproduce the qualitative trends found for bias fields induced by defect dipoles
by means of \emph{ab initio}-based simulations.

\bibliography{MyLibrary,library}% Produces the bibliography via BibTeX.
\bibliographystyle{apsrev4-1.bst} 

\end{document}